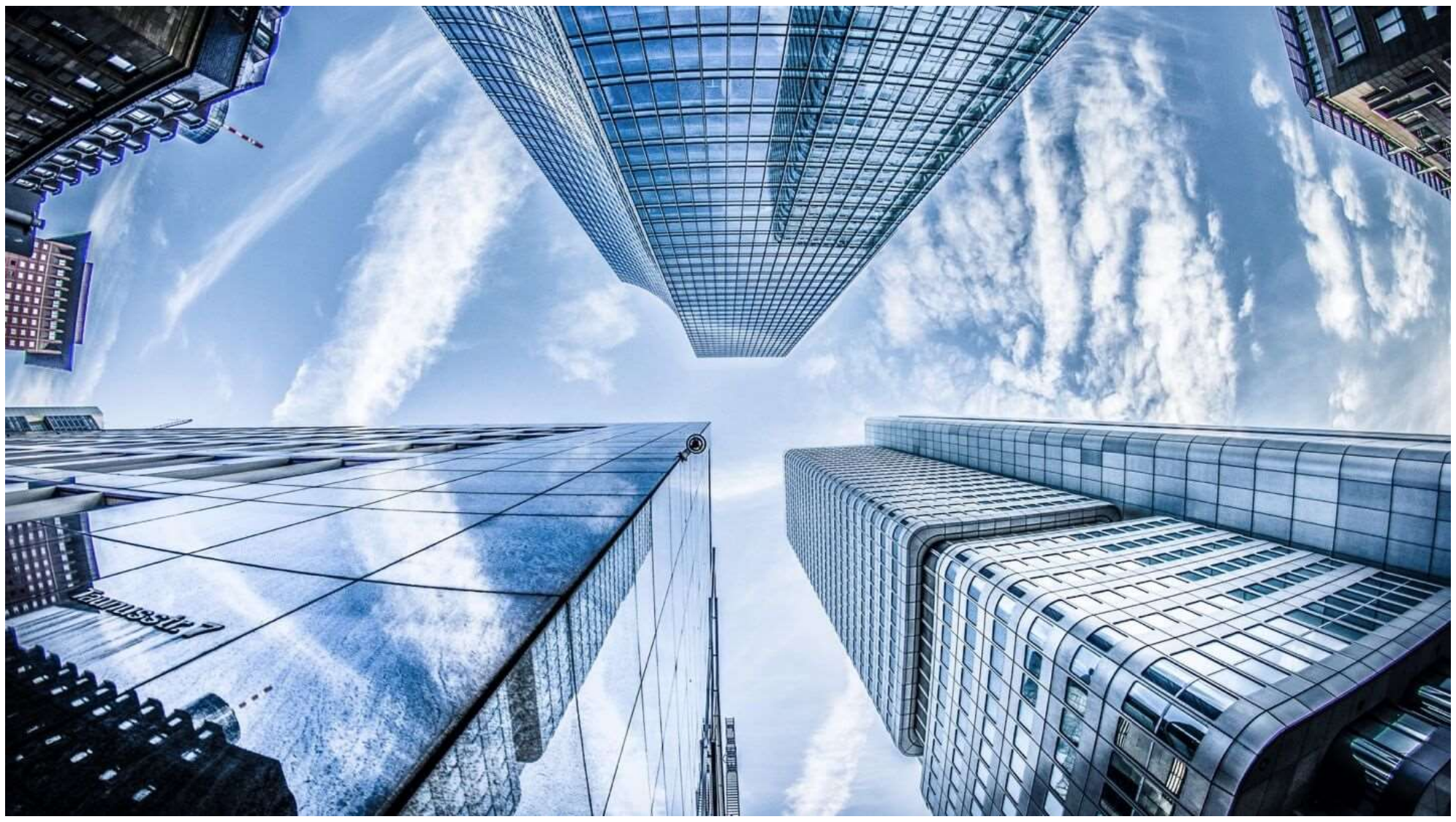

# Impermanent Loss in Uniswap v3

Stefan Loesch[†] stefan@topaze.blue

Nate Hindman nate@bancor.network

Nicholas Welch nick@bancor.network

Mark B. Richardson mark@bancor.network

## Table of Contents

† To whom correspondence should be directed.

## Abstract

Automated Market Makers ("AMMs") are autonomous smart contracts deployed on a blockchain that make markets between different assets that live on that chain. In this paper we are examining a specific class of AMMs called Constant Function Market Makers ("CFMMs") whose trading profile (ignoring fees) is determined by their *bonding curve*, eg $k = x * y$. This class of AMM suffers from what is commonly referred to as Impermanent Loss ("IL"), which we have previously identified as the Gamma component of the associated self-financing trading strategy and which is the risk that LP providers wager against potential fee earnings.

The recent Uniswap v3 release has popularized the concept of leveraged liquidity provision - wherein the trading range in which liquidity is provided is reduced and achieves a higher degree of capital efficiency through elimination of unused collateral. This leverage increases the fees earned, but it also increases the risk taken, ie the IL. Fee levels on Uniswap v3 are well publicized so, in this paper, we focus on calculating the IL.

We found that for the 17 pools we analyzed – covering 43% of TVL and chosen by size, composite tokens and data availability – total fees earned since inception until the cut-off date was $199.3m. We also found that the total IL suffered by LPs during this period was $260.1m, meaning that in aggregate those LPs would have been better off by $60.8m had they simply HODLd.

We then try to identify which class of users, if any, does better than others. Our first hypothesis was that active users who adjust their positions more frequently would outperform those that don't, but we could find no statistically meaningful evidence for that. We then looked at the time frame during which the position was active. We found that, with the exception of flash LPs who provided liquidity intra-block, all time segments lost. We did however note a trend that longer term LPs, whilst not making money after IL, did better than short term LPs: the IL for LP positions >1 month was 1.1x fees whilst, this for up to a day was 1.8x fees.We identified a number of areas for further research, notably, understanding whether there are medium-term LP strategies that consistently outperform HODLing, or whether the best performing strategy is simply related to "time in the market as opposed to timing the market" as the proverb goes.

## Introduction

Uniswap released its version 3 in March 2021, which popularized and industrialized the concept of leveraged or concentrated liquidity. Uniswap V3 is a clever reengineering of the leveraged liquidity concept, but it is not the seminal example; its predecessors include the stableswap invariant introduced by Curve Finance v1 and the amplified liquidity of Bancor v2.0. Impermanent loss (“IL”) in a concentrated liquidity framework is complex as it must contend with the leverage introduced, while also accounting for the IL outside of the range, which does not exist for traditional AMMs.

In this paper we extend the definition of IL to leveraged market makers, introducing the concepts of *minimal IL* and *actual IL*, and derive the relevant formulas for their computation. We then apply those concepts and formulas to Uniswap V3 where we analyze pools with uncorrelated pairs (eg stable coin pairs have been excluded) and a TVL greater than USD 10m, which at the time of this study accounted for roughly 43% of all TVL on Uniswap V3. The pools were chosen based first on TVL size and then excluded if the required external price data was unavailable.

Our core finding is that overall, and for almost all pools, both the minimal and actual IL surpass the fees earned during this period. In other words, the average liquidity provider (“LP”) in the Uniswap v3 ecosystem has been financially harmed by their choice of activities and would have been more profitable simply holding their assets ("HODLing”). Importantly, this conclusion appears broadly applicable; we have collected evidence that suggests both inexperienced retail users and sophisticated professionals experience a comparable struggle to turn a profit under this model, with the exception of “flash LPs” (aka just-in-time or “JIT” liquidity providers) who provide liquidity for a single block, to absorb fees from upcoming trades, then instantly remove their position. This cohort generated a modest $1.27m in fees while incurring zero IL.

Automated market makers (“AMMs”) – specifically constant-product AMMs have experienced exceptional growth in total value locked (“TVL”) and fee income since their introduction. The major issue with providing liquidity in AMMs is exposure to Impermanent Loss, which we will discuss in more detail in the next section. Several new AMM designs have appeared to address this critical pain point; Balancer[1] attempts to curtail IL with multi-asset pools and variable weights, Curve[2] has augmented the constant product function and primarily focusses its activities on stable assets to restrict IL, Thorchain[3] uses a variable fee structure, and Bancor v2.1[4,5] provides IL insurance at the protocol level.

Uniswap v2 was the most successful of all AMMs as far as volume and fees are concerned. Uniswap v3 was introduced with the publication[6] of its whitepaper in March 2021, and launched in May 2021 with a strong focus on “leveraged” or “concentrated” liquidity. The new model has attracted significant attention from both retail and professional users. Notably, Uniswap v3 has provided fertile ground for the creation of secondary DeFi protocols that seek to establish a reliable source of revenue generation from new strategies inspired by the leveraged liquidity paradigm.

Uniswap v3’s performance has been impressive since launch. Across analyzed pools, Uniswap generated over $108.5b in trading volume and a total of $199.3m in fee income from May 5th 2021 to September 20th 2021. However, what is often forgotten is that AMM fees do not come for free – they are a compensation for the risk, in particular the IL risk – that liquidity providers (“LPs”) are taking. While Uniswap V3 generated $199.3m in fees, during the same period the analyzed pools incurred, according to our calculations, over $260.1m in impermanent loss, leaving half (49.5 %) of liquidity providers with negative returns. To the best of our knowledge there has, to date, been no systematic study of the IL suffered by Uniswap v3 LPs, which is the focus of our work here.

This paper is divided into the following sections.

After this introduction we review the concept of AMMs and CFMMs, with a special focus on IL. We then extend the theoretical framework to the type of levered / concentrated AMM that Uniswap v3 represents. In Section IV, we provide some descriptive global and pool level statistics, allowing us to better understand the shape of the system.

Sections V and VI are the core sections of this paper: we first analyze the IL across the pools and compare it with the fee levels the pools offer to understand whether LPs derive economic benefit from their activity. In the last section (VI), we look at ROIs and other performance results including the percentage of LPs making vs. losing money and their mean returns.

# Review and Methodology

## Automated market makers

In May 2021, Uniswap launched their latest protocol version, Uniswap v3 which heavily emphasizes the concept of concentrated liquidity. Concentrated liquidity itself is not entirely new, but Uniswap v3 pushed the envelope a little further. In Uniswap v3, for each of the pools, the trading area is segmented into specific price buckets and liquidity providers can freely choose the range of buckets within which they want their liquidity to be active. Therefore, Uniswap v3 can achieve significantly lower slippage at the same level of liquidity provided and, on the not unreasonable assumption that trading volume follows slippage, also achieve a significant higher level of fees per unit of liquidity provided.

Those higher fees however do not come for free, and this is often ignored: leveraging liquidity not only amplifies the returns, but also the risks associated with those returns. In this case, the trade-off is that for increased fee levels, the liquidity providers must also be exposed to increased levels of IL. The purpose of this work is to dive into this subject and to analyze the IL that liquidity providers on Uv3 suffer.

## Impermanent Loss in a traditional AMM

Before we dive into the analysis of IL, we need to point out that this terminology is very unfortunate: the so-called "impermanent" loss is often anything but impermanent, especially in relation to coins that rapidly increase in value ("moon", using the common vernacular). An investment that would have appreciated 50x if held outside of an AMM is handicapped to a mere 10x if the coins are used for liquidity provision at an AMM. The fact that this IL would be reversed if ever the coin in question returns to its original value is of little comfort to the "unlucky" liquidity provider who was probably also a believer in that coin.

This is a convenient segue into our definition of IL, which we want to formally define as:

> Impermanent Loss is the difference between the value of the current fee adjusted liquidity position in an AMM, and the HODL value of the position that was originally contributed.

The reason the current position needs to be adjusted is that AMM positions suffer gas costs and earn fees, and the impact of those needs to be excluded if we are to discriminate the IL. Also, we want to point out at this stage that this definition becomes more involved in Uv3 as the reference HODL position is not easy to identify in the presence of multiple liquidity transactions.

We have discussed the topic of IL, and its relation to option pricing, in a previous paper[5], and we refer the reader to this paper for a more in-depth discussion. Here we only want to recall the key results.

A traditional AMM – also referred to as constant product AMM – relies on the following characteristic curve

$$k = m \cdot n$$

where $k$ is the pool constant, and $n$ and $m$ are the pool constituents, measured in their own numeraire. In other word, if a pool contains 2 ETH and \$2,000 then the constant is $k = 2 * 2{,}000 = 4{,}000$. The AMM then offers any trade that maintains the pool constant, well, constant; in other words, and ignoring fees, the AMM is indifferent holding 2 ETH and \$2,000 USD, or 1 ETH and \$4,000, or 8 ETH and \$500 and will engage into any trade that leaves it in one of those end states. It is worth mentioning that if k is constant so is any function $f(k)$. Often it is preferred using $k = \sqrt{m \cdot n}$ instead of the above, but as far as the AMM is concerned those two are the same.

Arguably the most important insight with respect to traditional AMMs is the following.

> In a traditional AMM, the aggregate dollar value of the two assets in a fully arbitraged AMM is always equal.

We are referring to the *dollar value* in the statement above, but of course if the value in dollar is equal so is the value in Euro, in ETH or in any other numeraire chosen. A key corollary to this – and the ultimate source of the IL – is the following statement:

> An AMM always sells the outperforming asset and buys the underperforming asset.

It is easy to see that this leads to a loss against HODL: the initial pool contribution was equal in value. Then the pool keeps selling the outperforming asset as it goes up. In other words, during a rally, instead of HODLing, the pool sells the outperforming asset into the rally, meaning it can no longer profit from subsequent price increases. That, in a nutshell, is the source of IL.

If you are familiar with option pricing you might recognize this as a *positive Gamma strategy*. In a range trading market this strategy allows you to make money by *buying low* (you buy when an asset underperforms) *and selling high* (you sell when it outperforms), and you should therefore earn the option premium "Theta" on your "Gamma". It turns out however that here this is not the case, and the reason is that AMMs are not active traders but instead rely on arbitrageurs. Ignoring fees, a standard AMM offers an arbitrageur to execute the trade at the mid-point (geometric average) of the pool-implied price before and after the trade.

It is worth playing this through to understand the nature of IL: we assume that the pool-implied price is currently at 100, and the price moves to 110. In this case the pool offers to trade at $\sqrt{100 \times 110} \sim 104.8$, leaving a profit of 5.2 for the arbitrageur. If market then goes back to 100, the pool again allows to trade at $\sqrt{110 \times 100} \sim 104.8$, but this time in the other direction. In other words: the pool has traded twice at 104.8, in opposite directions, and is therefore exactly where it started. This is the reason why this is referred to as "impermanent" loss – it disappears when markets return to their initial levels. However, the pool leaked significant value: it sold at 104.8 when it could have sold at 110, and it bought at 104.8 when it could have bought at 100. Overall, the pool could have made 10 from this move but, ignoring fees, it paid those 10 to the arbitrageur.

In order to calculate IL, it is easiest to place ourselves in the numeraire on one of the assets – say USDX for an ETHUSDX pool. We define $x_t$ as the *exchange ratio* of the two assets, ie the ratio between the two prices. In the case of ETHUSDX and USDX as numeraire is simply the ETH price expressed in USDX. We also normalize it to $x_0 = 1$, so technically our $x_t$ is the change in the price ratio if we are being precise. In this case the value of the HODL portfolio that invests 1 unit of the numeraire into each of the assets is

$$HODL(x) = 1 + x$$

Again, to be precise, we should point out that those numbers are dimensionless and need to be multiplied with the initial portfolio notional for scale. It is a well-known result[5] that the value of the AMM portfolio in the above case is

$$AMM(x) = 2\sqrt{x}$$

For the IL this gives us

$$IL_t = \frac{AMM_t - HODL_t}{HODL_0} = \frac{1}{2}\left(2\sqrt{x_t} - x_t - 1\right)$$

As discussed above, this number is dimensionless – it is simply a percentage applied to the original portfolio notional.

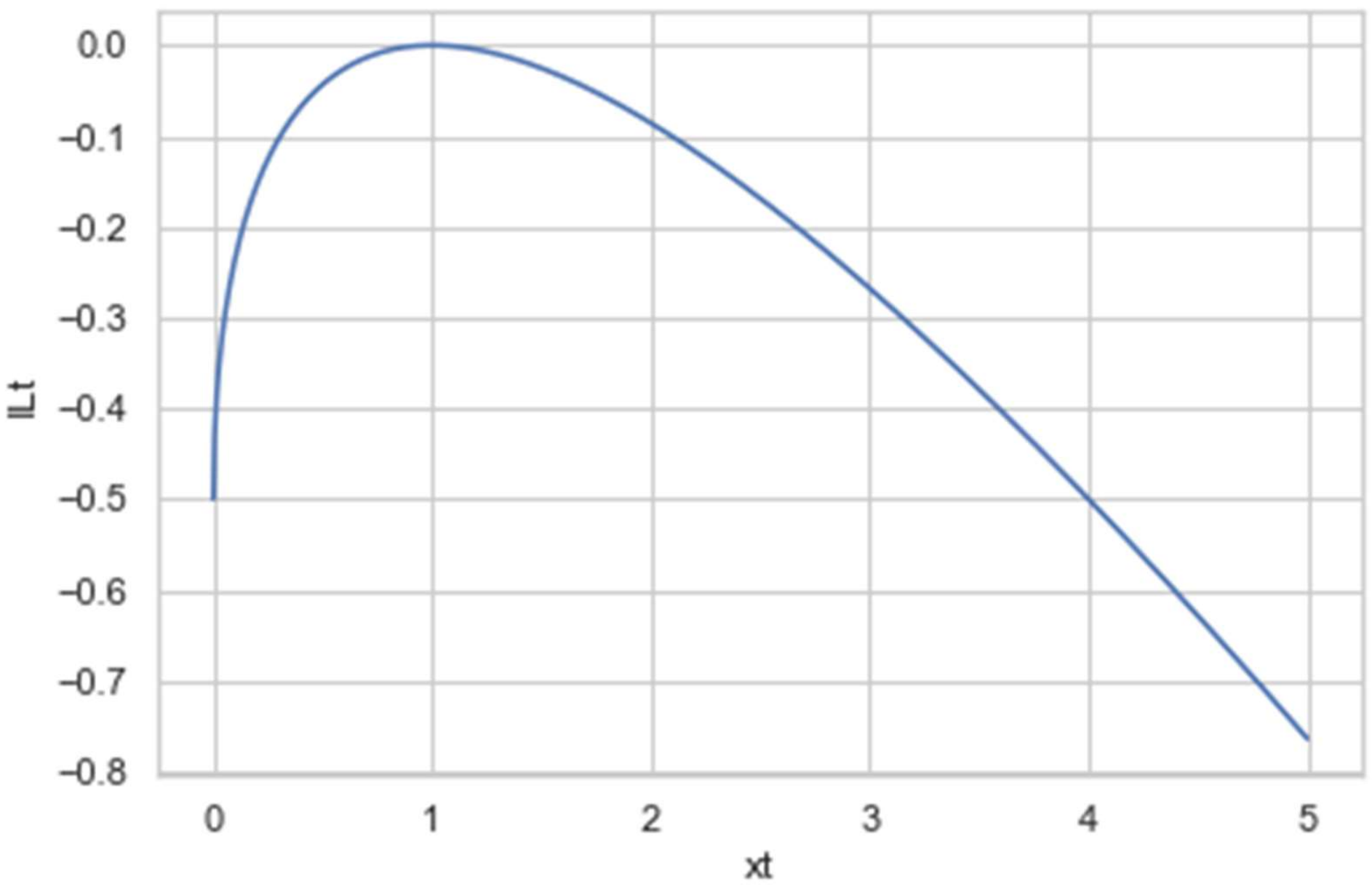

The IL on the upside is unlimited whilst on the downside it seems limited, but this is a consequence of choosing a numeraire. This is easiest to see with an example: if the pool is ETHUSD and ETH appreciates 100×, then in USD terms the pool goes 10x while HODL goes 50× (only 1/2 the HODL funds are in ETH). The IL is therefore 40× the initial investment, and we see that ultimately IL measured in USD is not limited in the ETH upside. On the ETH downside however, we will never be able to lose more than the initial investment, so the maximum IL *on the ETH downside, measured in USDX* is 100%. However, measured in ETH this loss is substantial as ETH goes to zero. We can just flip around that *if we measure everything in ETH* then the ETH downside (aka "USD mooning") has unlimited IL if measured in ETH.

Impermanent loss in Uniswap v3

Introduction

The Uniswap v3 formulas are somewhat complex. We have made an effort to organize the theory in such a way as to demonstrate that, in fact, the underlying principle is relatively familiar.

The key insight by the Uniswap developers is that we can limit the trading range of an AMM by introducing a hard stop. Suppose the initial exchange ratio $x$ was 100, and we simply force the AMM to stop trading if $x$ is outside the range 80 ... 120. We know that at 80 (20% down) the IL is $\frac{1}{2}(1 + 0.8 - \sqrt{0.8}) \sim 0.56\%$, and on the upside the IL is $\frac{1}{2}(1 + 1.2 - \sqrt{1.2}) \sim 0.46\%$. We also know that anywhere in the range, and therefore also at the boundaries, the dollar value in each of the constituent assets is the same. So, on the lower end, there are 80 units of the risk asset for each unit of numeraire, and at the upper end there are 120. We also know that below 80 and above 120 the AMM simply stops trading, so from this point onwards the portfolio composition (but not its value!) remains the same regardless of price changes.

Looking at this from a traditional AMM lens this limits our IL to about 0.5%. We however now have a positional mismatch, and as we will see below this is the single biggest problem to define a meaningful and intuitive IL measure for range-80 ... 120 AMMs. The issue is simple: we have seen that between 80…120 our IL was well inside 0.6%. So far so good, but what about outside this range? Our initial portfolio had 100 units of the numeraire asset for each unit of the risk asset. Below the lower bound the portfolio is static, but it is static at 80 units of numeraire per risk asset. In other words: the AMM is more exposed to the risk asset than our HODL portfolio (the AMM buys low, sells high) and therefore will suffer more if the risk asset falls further.

This directional loss is a real loss, but it is structurally different from the IL we're used to looking at so we are making an important distinction here:

**Minimum IL.** The minimum IL is the IL that is incurred while the asset was in range; this is not avoidable even by the most sophisticated traders – as long as they want to earn fees, that is the risk they are facing.

**Out-of-range IL.** The out-of-range IL is the IL incurred when the asset was outside the range; it is simply the difference in performance between the initial HODL position, and the frozen position at the end of the range that is overweight in the underperforming asset. Out of range IL can be avoided without fee loss by withdrawing the position after it goes out of range and rebalancing the portfolio into the initial HODL composition – or into any directional position that the LP prefers at this stage

**Actual IL.** The *actual IL* is the sum of the minimum IL and the out-of-range IL

We considered the *actual IL* as the most important IL in our analysis because we wanted to look at how real LPs fare on Uniswap v3. This means we need to include all incurred losses, even if they would have been avoidable by a more sophisticated investor who has the wherewithal to rebalance their position if it ever goes out of range.

So far, we have used the restriction of the trading range to 80…120 to derisk our traditional AMM. However, there is a second angle to this: because we limited the trading range there is a limit to how far the amount of constituent assets in the pool can fall. On the downside the AMM buys the risk asset and sells the numeraire asset, therefore the absolute minimum holdings of numeraire asset are in our case at an exchange ratio of 80. We know the value of the portfolio at 80, it is $200\sqrt{0.8}$, and we know that 50% of this is in the numeraire. In other words, the minimum numeraire holdings in the 80…120 AMM are $100\sqrt{0.8} \sim 89$. Instead of contributing \$100 worth of the numeraire we can contribute \$11 because this is the maximum we can ever lose.

> *This is the basic principle behind a levered AMM: we contributed only 11 units of the numeraire asset, but our AMM behaves as if it had 100 units worth of liquidity. The fact that we start at 80 gives us about 9x leverage on the amount of numeraire asset contributed.*

On the upside the calculation is similar, we just have to take into account that we now have numeraire effects. The value of the AMM portfolio at 120 is $200\sqrt{1.2}$ and again half of it is in the numeraire, so we have $100\sqrt{1.2} \sim 109.5$ units of numeraire invested in the risk asset. We also know that, at the boundary, 1 unit of the risk asset is worth 120, so in units of the risk asset our holding are $\frac{109.5}{120} = 0.91$. In other words, instead of starting with 1 unit of the risk asset we can start with 0.09 units of the risk asset, yielding 11x leverage.

Note that the upside and downside leverage numbers are not the same, and also the value of the assets contributed is no longer the same either: we need 11 units of the numeraire, and 0.09 assets of the risk asset which, at 100, corresponds to a value of 9. This is because our range is asymmetric when looked at it through a multiplicative lens. We can see this effect even more if we'd used a clearly asymmetric range like say 80…140. In this case what happens at the lower limit remains unchanged, but at the upper limit we have $\frac{100\sqrt{1.4}}{140} \sim 0.85$. In this case we still need 11 units of the numeraire but, because of the wider range at the upside, we need 15 numeraire units worth of the risk asset at the outset. Contrary, if we restrict our upside to 110 we get $\frac{100\sqrt{1.1}}{110} \sim 0.95$, so we only need to use 5 units worth of the risk asset.

Now the fact that we need to use less collateral is advantageous from one point of view. If we assumed that there is a limit to the amount of collateral we can contribute, a levered AMM allows us to run a bigger virtual pool with lower slippage, and therefore hopefully attract more trading and therefore earn more fees. The downside however is that this magnification also magnifies the IL. To give rough numbers for the 80 ... 120 range, the initial IL was 0.5% of the unlevered pool notional. Our maximum leverage in this range is about 10×, and this means our max IL suddenly increases by the same factor, to 5%. For more pronounced asymmetric ranges we need to distinguish between upside and downside IL.

Uniswap v3 portfolio value and composition

The value of a Uniswap v3 AMM portfolio is more complex, and for example Lambert[7] has a good overview of how to arrive at it. Using a slightly different notation to Lambert we find that the portfolio value $AMM(x)$ of a Uniswap v3 position is

$$AMM(x) = n_0 \cdot x \ \ \forall x < x_0$$

$$AMM(x) = n_0 \cdot \left( \sqrt{x_0 x_1} \cdot \frac{\sqrt{x} - \sqrt{x_0}}{\sqrt{x_1} - \sqrt{x_0}} + \sqrt{x_0 x} \cdot \frac{\sqrt{x_1} - \sqrt{x}}{\sqrt{x_1} - \sqrt{x_0}} \right) \ \ \forall x_0 < x < x_1$$

$$AMM(x) = n_0 \cdot \sqrt{x_0 x_1} \ \ \forall x > x_1$$

Here $x$ is as usual the price ratio of the two assets in the pool, $x_0$ and $x_1$ are the lower and upper bounds of the range where liquidity is provided, and $n_0$ is an overall portfolio notional factor. We have plotted this function for 3 different ranges below, with the lower bound of the range being x=70, 80, 90 respectively, and the upper bound chosen so that the range is symmetric around *x=100* in multiplicative terms.

Value Uniswap v3 portfolio at different ranges

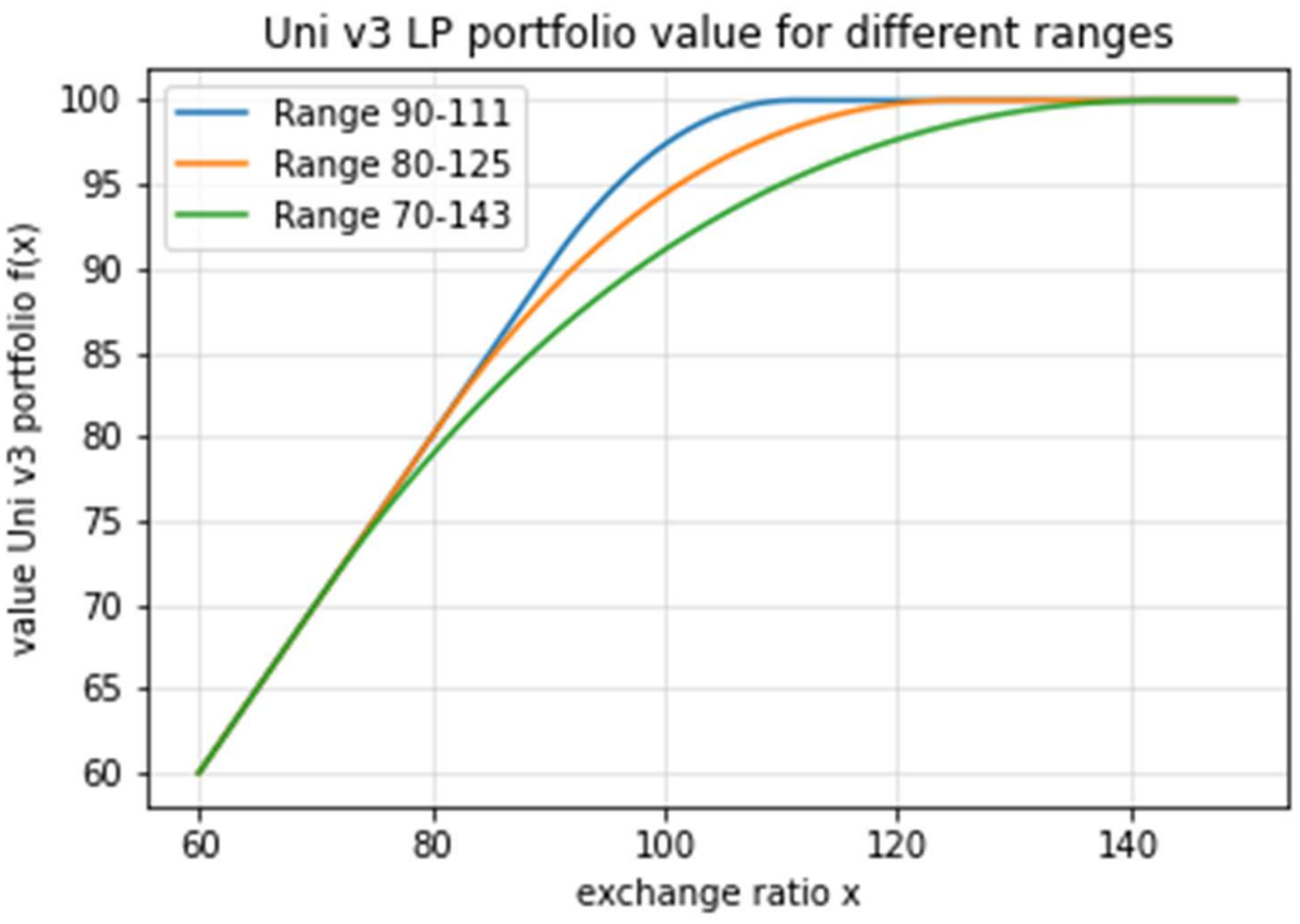

We see that for $x < x_0$ the chart is linear in $x$, meaning that the position is entirely vested in the risk asset as we would expect it to be at the risk-asset downside. For $x > x_1$ the curve is constant, meaning that the position is entirely vested in the numeraire asset, again as we'd expect at the risk-asset upside.

The normalization of the value function is such that at the risk-asset downside, exactly 1 unit of the risk asset is held. On the upside this will be converted into the numeraire at the (geometric) average price of the range which is $\bar{x} = \sqrt{x_0 x_1}$. That conversion ratio is to be expected: we know that when a traditional AMM adjusts from $x = x_0$ to $x = x_1$ then it does so at an exchange ratio of $\bar{x} = \sqrt{x_0 x_1}$, and Uniswap v3 behaves between $x_0 \dots x_1$, like a traditional AMM, so its behavior has to be consistent with that.

Note that we can find the $AMM(x) = 2\sqrt{x}$ of a traditional AMM by first setting $n_0 = 1/\sqrt{x_0}$. This is just a portfolio-notional consideration because the portfolio size normalization inherent at $n_0 = const$ does not make sense at the limit $x_0 \to 0$. Then we can set $x_0 = 0.$, and finally we can take the limit $x_1 \to \infty$.

The holdings of the *risk asset* corresponding to the aforementioned valuation formula are

$$N_r(x) = n_0 \cdot 1 \;\; \forall x < x_0$$

$$N_r(x) = n_0 \cdot \sqrt{\frac{x_0}{x}} \cdot \frac{\sqrt{x_1} - \sqrt{x}}{\sqrt{x_1} - \sqrt{x_0}} \;\; \forall x_0 < x < x_1$$

$$N_r(x) = 0 \;\; \forall x > x_1$$

and those for the *numeraire asset* are

$$N_n(x) = 0 \;\; \forall x < x_0$$

$$N_n(x) = n_0 \cdot \sqrt{x_0 x_1} \cdot \frac{\sqrt{x} - \sqrt{x_0}}{\sqrt{x_1} - \sqrt{x_0}} \;\; \forall x_0 < x < x_1$$

$$N_n(x) = n_0 \cdot \sqrt{x_0 x_1} \;\; \forall x > x_1$$

where $n_0$ is an overall portfolio notional factor.

As a reference, here is the comparison between the value of a Uniswap v3 vs a Uniswap v2 liquidity portfolio

Value Uniswap v2 vs v3 portfolio

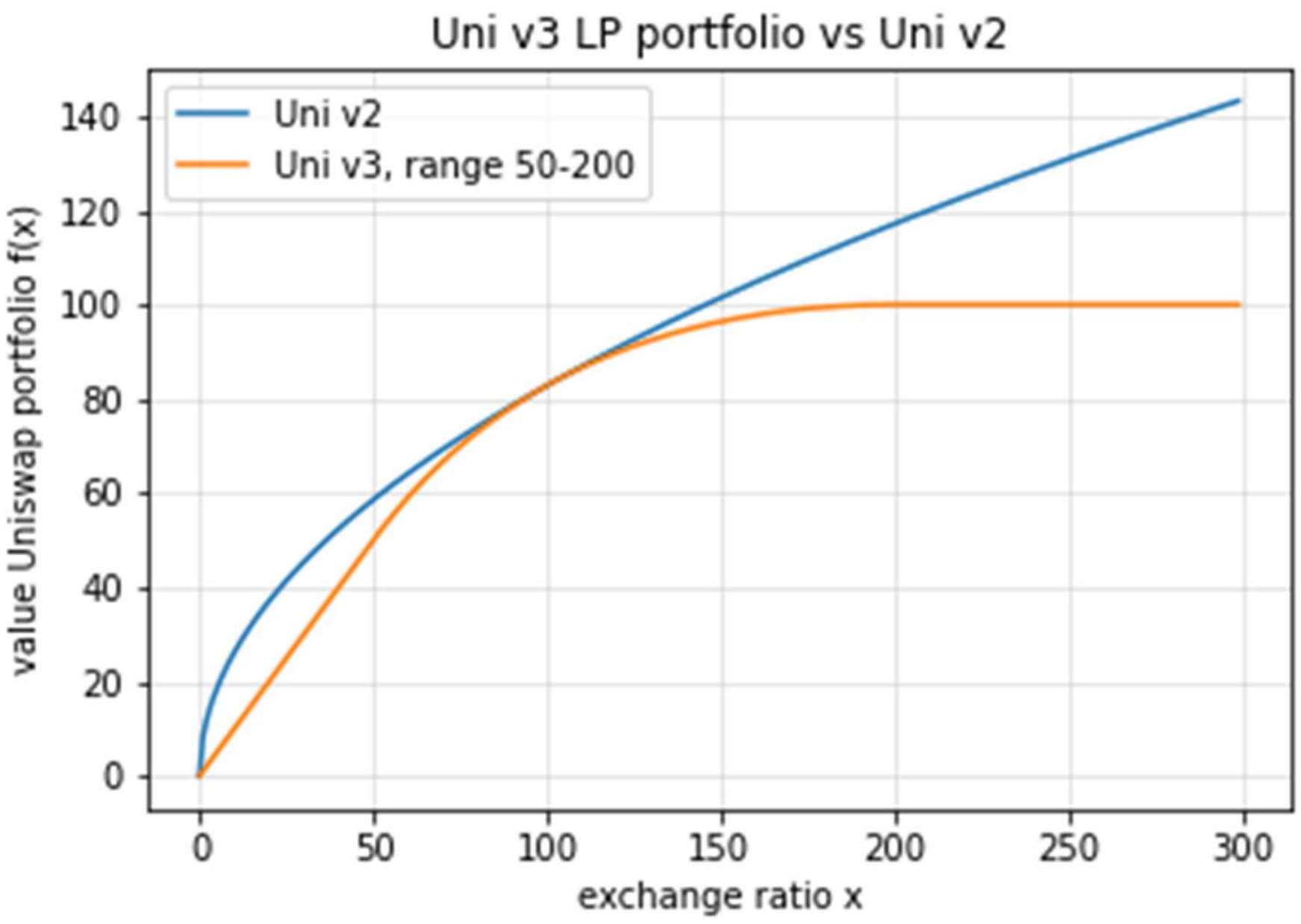

We see that while Uniswap v2 offers some participation in the upside of the risk asset (the LP portfolio value grows like $\sqrt{x}$ when the risk asset grows like $x$), the upside of Uniswap

v3 is flat over and beyond the upper end of the range. Likewise on the downside, the value of the Uniswap v3 portfolio descends significantly quicker than that of the Uniswap v2 portfolio ($\sqrt{x}$ for v2 vs $x$ for v3), even though ultimately, at $x = 0$, they coincide at a zero value.

Uniswap v3 Impermanent Loss

Impermanent Loss is defined as the difference between the HODL performance of the contributed assets, and the performance of the corresponding AMM portfolio. In a traditional AMM assets are always contributed in equal value, therefore the HODL value of a portfolio in terms of the numeraire asset is proportional to $1 + x$, where the 1 represents one unit of the numeraire asset and $x$ represents one unit of the risk asset, and the exchange ratio $x$ is normalised to $x(t = 0) = 1$.

For Uniswap v3 we need to modify this formula in that our reference portfolio now consists of $N_n(x)$ units of the numeraire asset, and $N_r(x)$ units of the risk asset, with a suitable normalization choice for $n_0$. This of course is straight forward.

We also need to take into account that liquidity providers can add or remove liquidity over time, and that this will generally happen at $x(t) \neq x(0)$ and therefore at a different asset composition. The way we have dealt with this issue in that we consider that every time liquidity is added or removed, we consider that this as a novation of the position. In other words, we assume that at every change in liquidity the entire position is closed, and a new position corresponding the post-liquidity holdings is opened. Moreover, for IL calculation purposes we assumed that all open positions are closed out at the point of calculation.

To calculate the IL of one of those imputed positions we place ourselves in the numeraire of one of the participating tokens. It is easy to verify that for calculating the percentage IL the choice of numeraire does not matter as long as the numeraire is one of the two tokens involved or a linear combination thereof. However, it is also easy to see that in general the choice of numeraire does matter. As an example, we consider an ETH BTC pool where ETH appreciates by 20% against the USD and BTC appreciates by 10%. Whilst this position has a (small) IL when considered in either BTC or ETH, in a USD numeraire the IL becomes a gain instead of a loss because the effect of the average price rise outweighs the impact of the price divergence.

Once we have the percentage IL number, we convert this into a USD number by multiplying with the spot exchange rate at the time the imputed position was closed for further aggregation. We note that the choice of numeraire here is not an innocent choice in that it does influence outcome in a non-trivial manner when exchange rates and asset prices are changing significantly. However, it does guarantee that IL is always a loss and never a gain, an outcome that naive IL aggregation methods cannot guarantee, and we have convinced ourselves that the variation introduced by the choice of numeraire does in this case not alter our findings substantially.

## Descriptive Pool Level Statistics

In this section we focus on a number of statistics to set the scene, and to lay the groundwork for what we will discuss later. These numbers are by and large well known, and typically accessible either via the Uniswap v3 user interface, or via data services like Dune Analytics.

### Introduction and methodology

Our entire analysis is based on a dataset pulled on September 20th, 2021. It should be mostly self-consistent. However, as the process of pulling the data takes a number of hours some inconsistencies cannot be avoided. Also, we often translate numbers into USD. In order to do this, we use hourly dollar rates from querying Dune Analytics' *prices.usd* and we match all events to the closest hourly data.

Uniswap v3 has 3 types of pools that differ in fee levels and tick spacing. The idea is that high fee pools serve highly volatile pairs, and therefore a bigger grid spacing is appropriate, and vice versa. However, there is no requirement to choose any specific pool for any specific pair, and for many of the pairs there are multiple pools available competing with each other for liquidity and trading.

When naming a pool, we follow Uniswap conventions by starting with the two constituent assets and then to appended to the name of the pool is the fee level of that pool. The fee level can only have three distinct values: 10000 for a 1% fee pool, 3000 for a 0.3% fee pool, and 500 for a 0.05% pool.

For example, the USDCWETH3000 pool is the pool trading USDC against wrapped ETH, at a fee level of 30bp. Contrary to ISO currency descriptions, within the Uniswap UI there is no consistency between the way the pool exchange ratios are quoted and the order in which they appear in the pool name. We generally follow the convention that if one of the constituents is a stable coin, we use it as numeraire. Otherwise, we prefer exchange ratios >1 to those <1.

As already alluded to above, the perennial problem when aggregating data across many different assets that are highly volatile against each other is that the choice of numeraire is not an innocent choice. This can even affect numbers as straightforward as aggregate trade volume and fee levels.

To give an example, we assume two trading periods with 100m worth of TKN trades each, and associated fees of 3000 TKN in each period. Furthermore, we assume that TKN was at 1,000 USD in period 1, and at 2,000 USD in period 2. Converting volume and fees into USD in the respective periods yields 100b USD of trading in period 1 and 200b USD in period 2, with the corresponding fee income of 3m and 6m USD respectively, and a total volume of 300b USD and 9m USD in fees. Converting all numbers at the end however yields 400b USD in volume and 12m USD in fees. There is no single right answer to the above question. We often prefer to convert numbers into USD when they accrue as opposed to at the end, as the change due to changes in the exchange rate can lead to perplexing results. However, this also depends on the specific analysis undertaken, and there is no hard and fast rule.

TVL Statistics

In this section here we look at the Total Value Locked in the different pools, both at the last day of the analysis, and averaged over time, with exchange rate conversion at the end date.

Latest TVL per pool

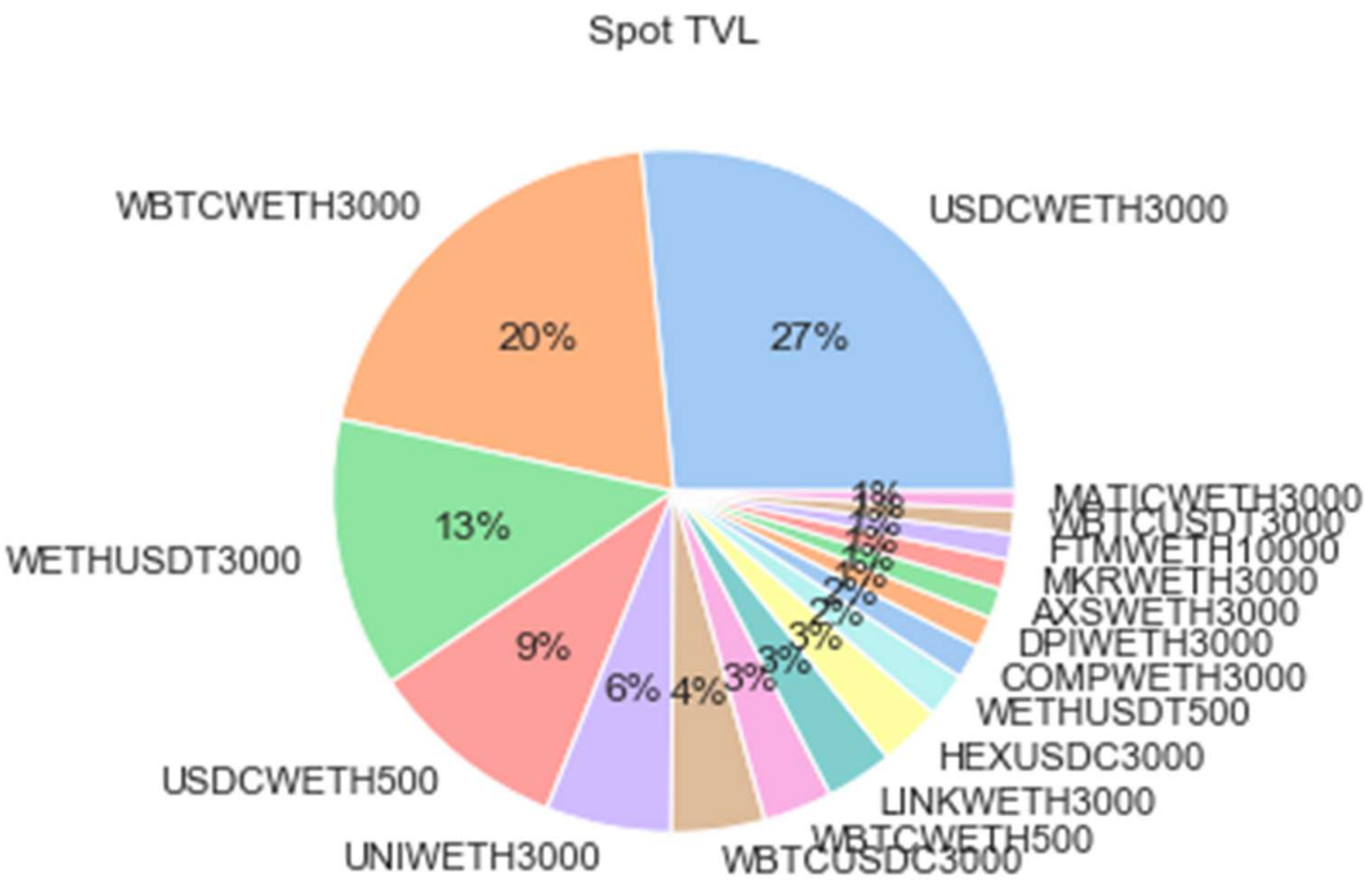

Average TVL per pool

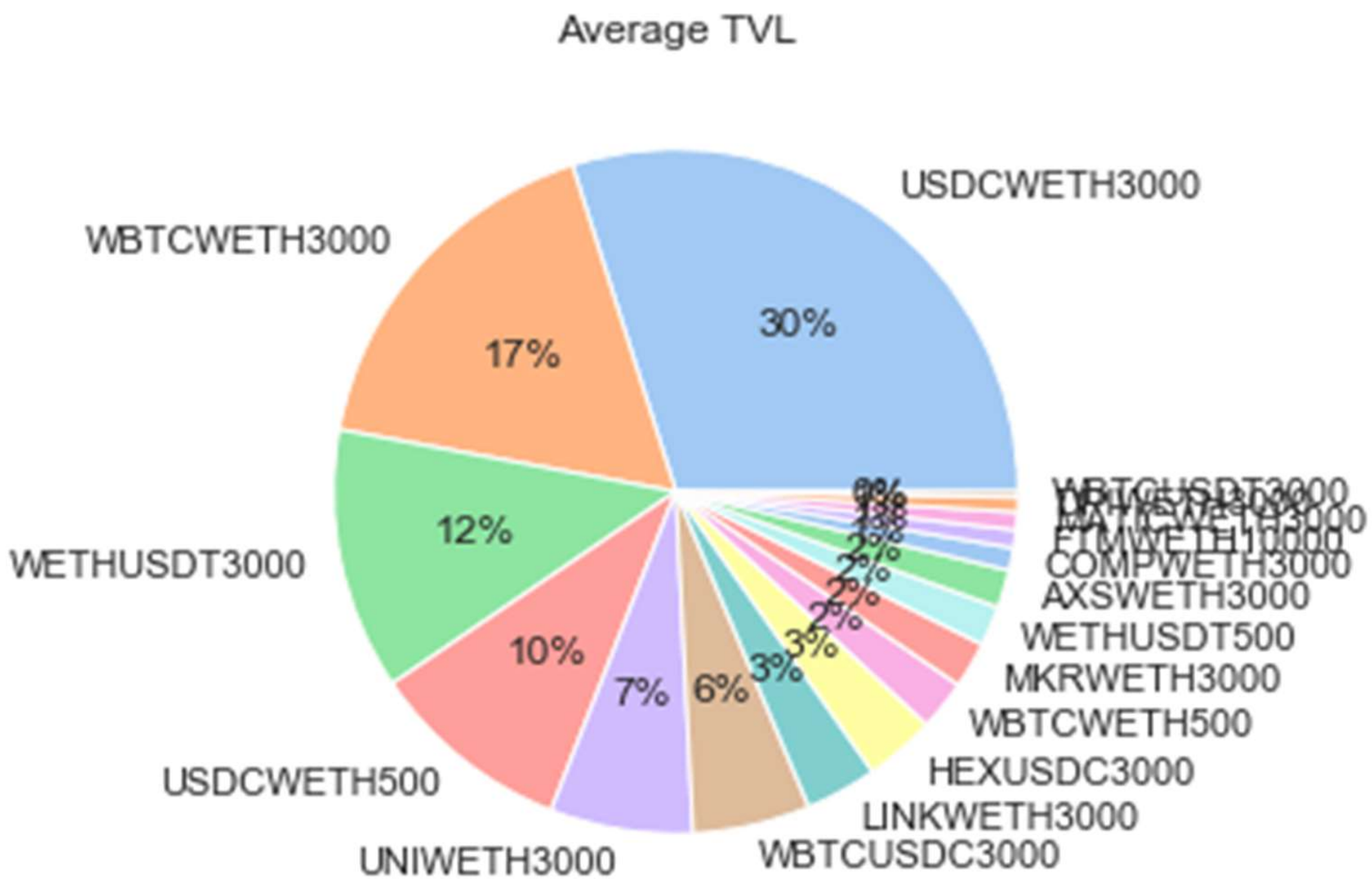

We firstly notice that the choice "average vs latest numbers" does make a difference in some cases, but substantially the charts are very similar. Our key takeaway from those charts is that Uniswap v3 is very much geared to the biggest pools – the biggest pool alone (USDCWETH3000) has 25-30% of all TVL, and the two biggest ones together are close to 50%. Moreover over 75% of the TVL is in pools that trade ETH or BTC against a USD stable coin or against each other. The biggest pool that does not fit into this category is UNIWETH with 6-7% of all TVL, depending on the measure, and from there the pool size descends quickly.

Fee and trade volume statistics

Having looked at TVL – the key driver of Impermanent Loss – we now look at the other side of the equation, the trade volume, and more importantly the associated fees. Optically the fees pie chart looks very similar to the TVL pie charts, with the notable observation that WETHUSDT3000 and WBTCWETH3000 switch places – the share of fees of the stable coin pools is substantially higher than its TVL share when compared to the ETH WBTC pool.

Aggregate fees per pool (from trade volume)

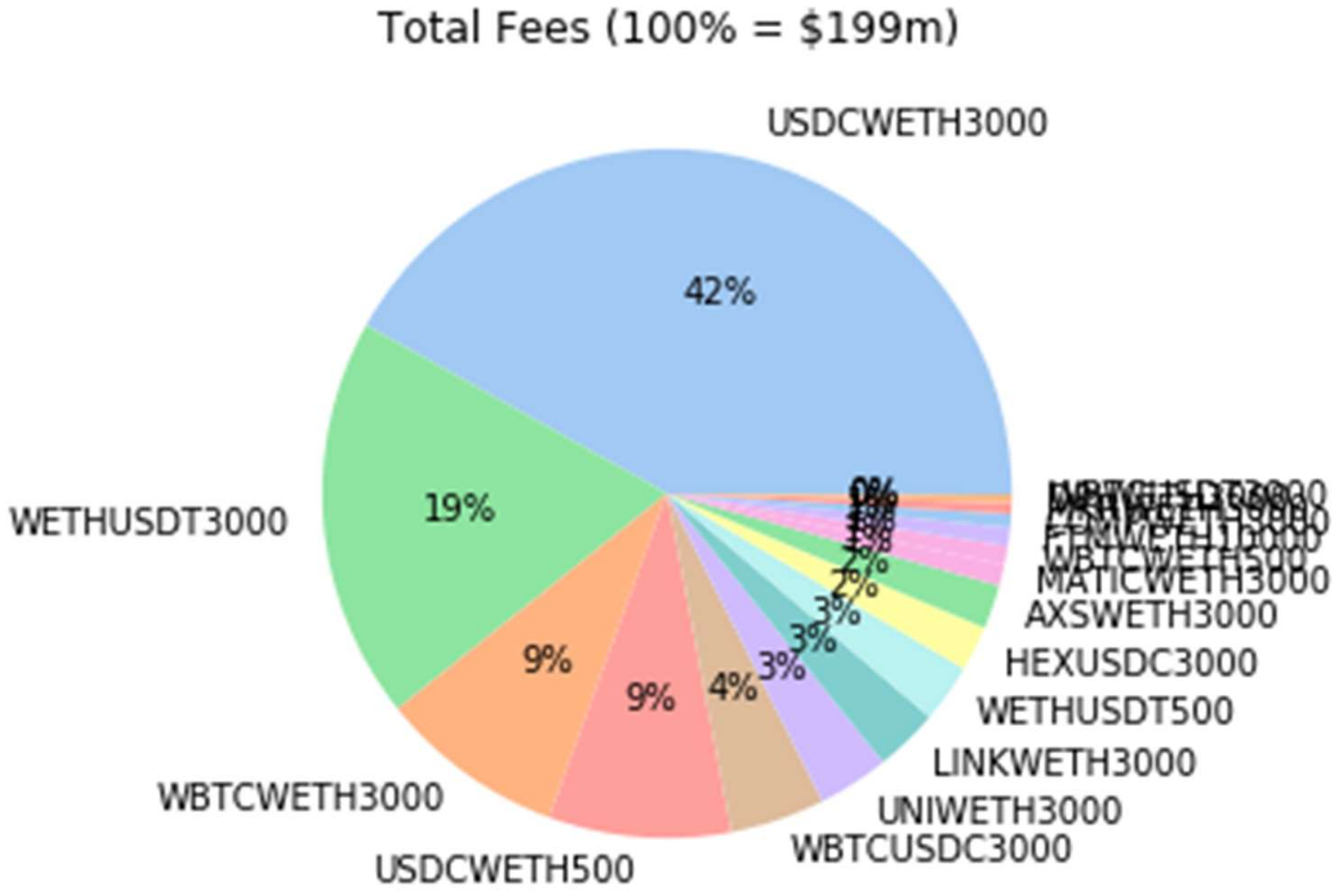

The other notable observation here is that fees in the 5bp USDCWETH500 pool are only about 25% of those in its 30bp brethren, despite the 5bp TVL being only about 1/3 of the 30bp TVL, and the fee level being only 1/6.

Trade volume per pool

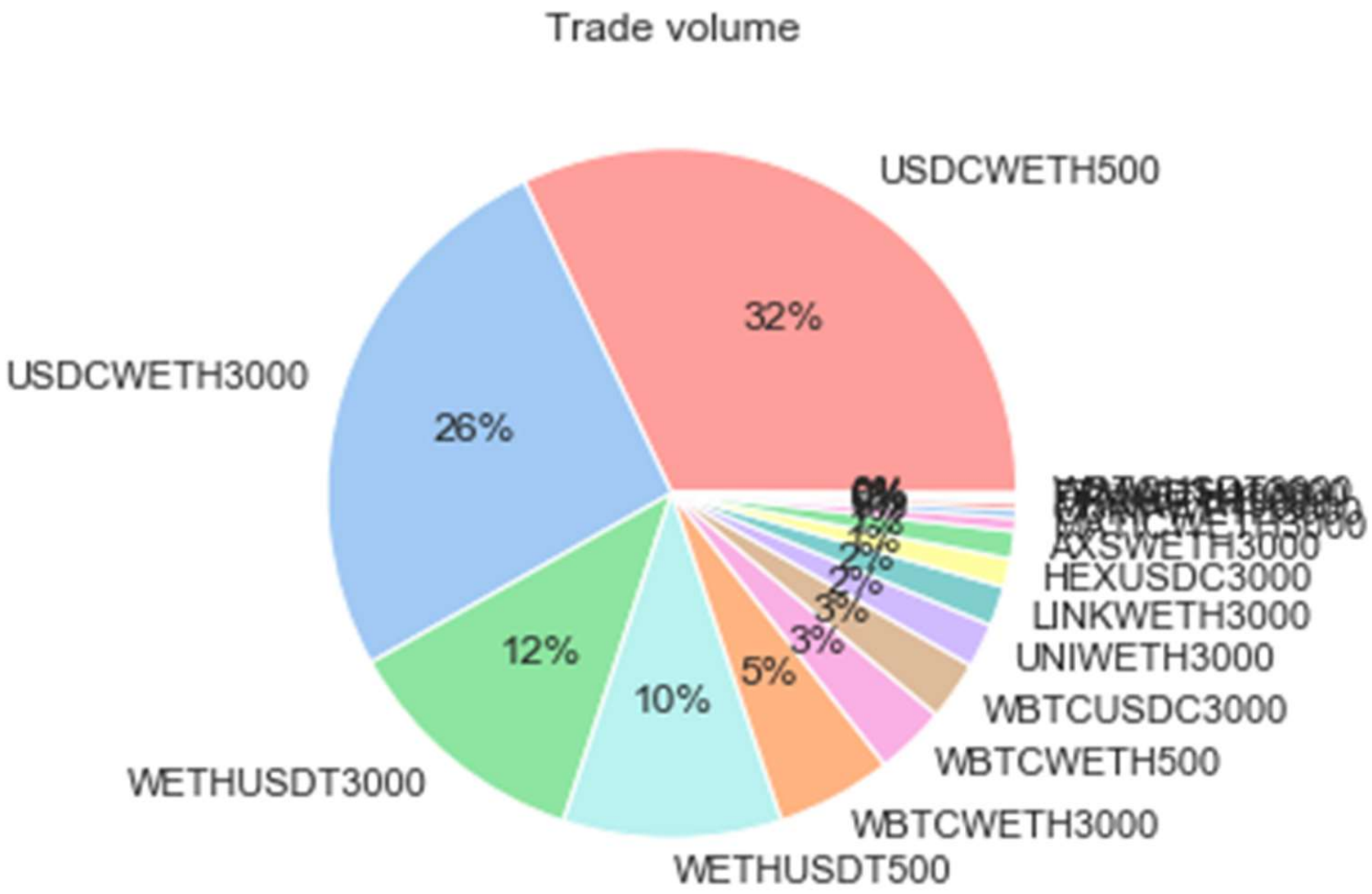

Positions and address statistics

We will now look at unique positions, and at unique addresses. It should be noted that positions can be modified over time and as we've discussed above in the IL calculation, we consider every modification its own *imputed* position. This however is not the case here – positions in this analysis are simply Uniswap v3 NFTs, regardless of the number of times they are modified.

Unique addresses are consolidated on a per-pool but not on a cross-pool basis. We have not correlated different addresses operated by a single owner, so it is likely to be bigger than the number of LPs, and the biggest / most active addresses may not necessarily correspond to be biggest / most active LPs.

Unique addresses

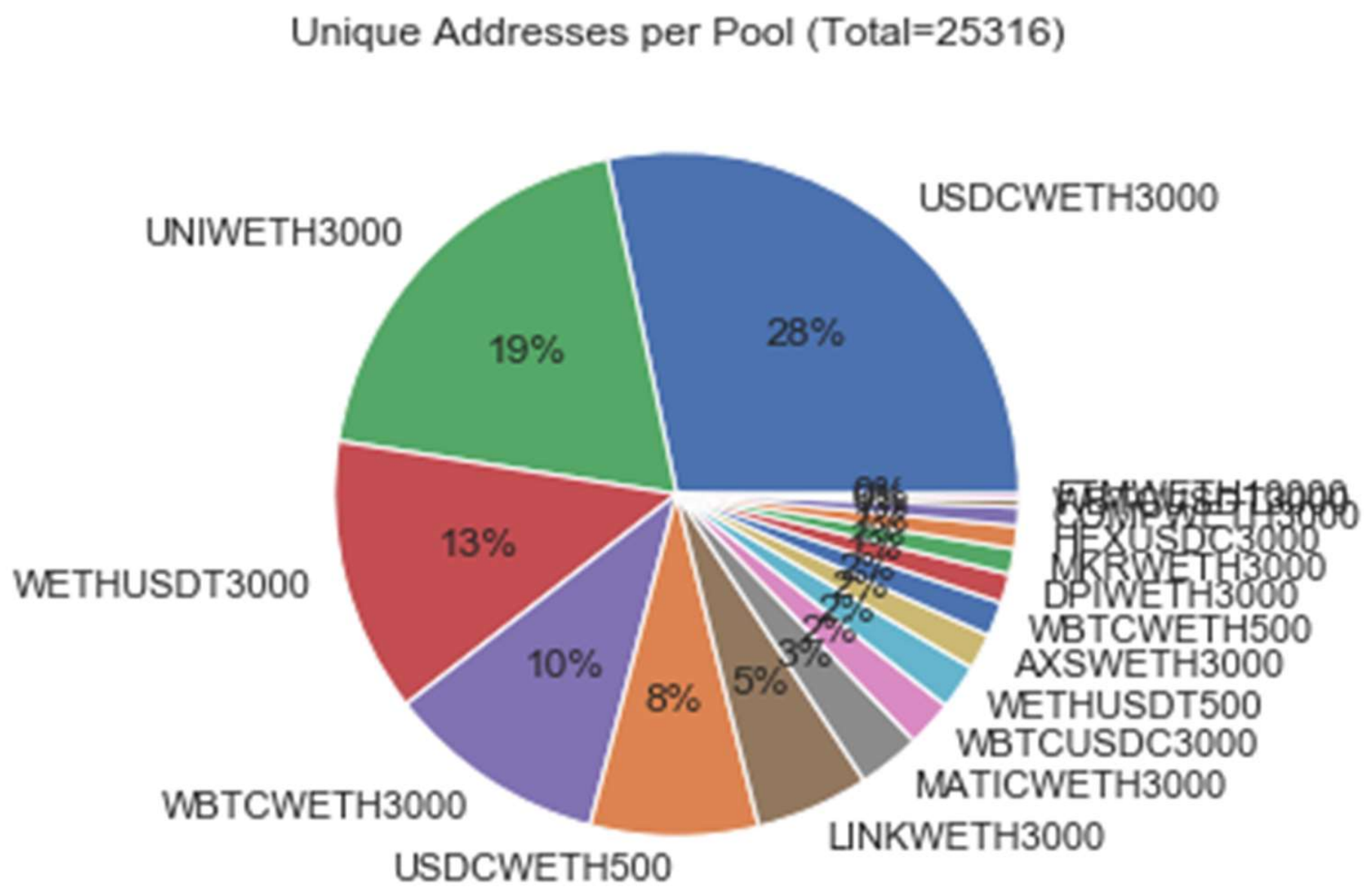

Positions

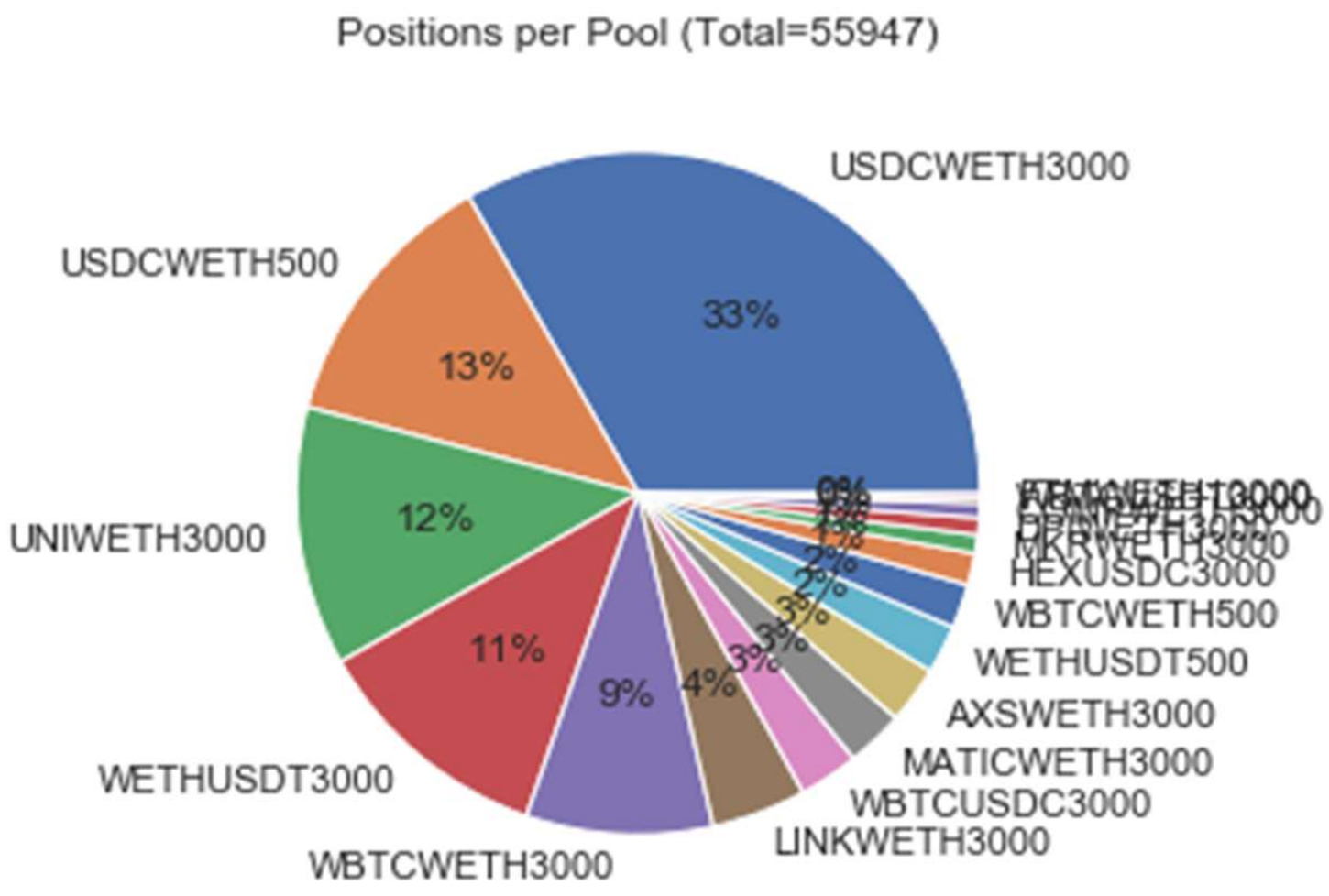

Positions and addresses

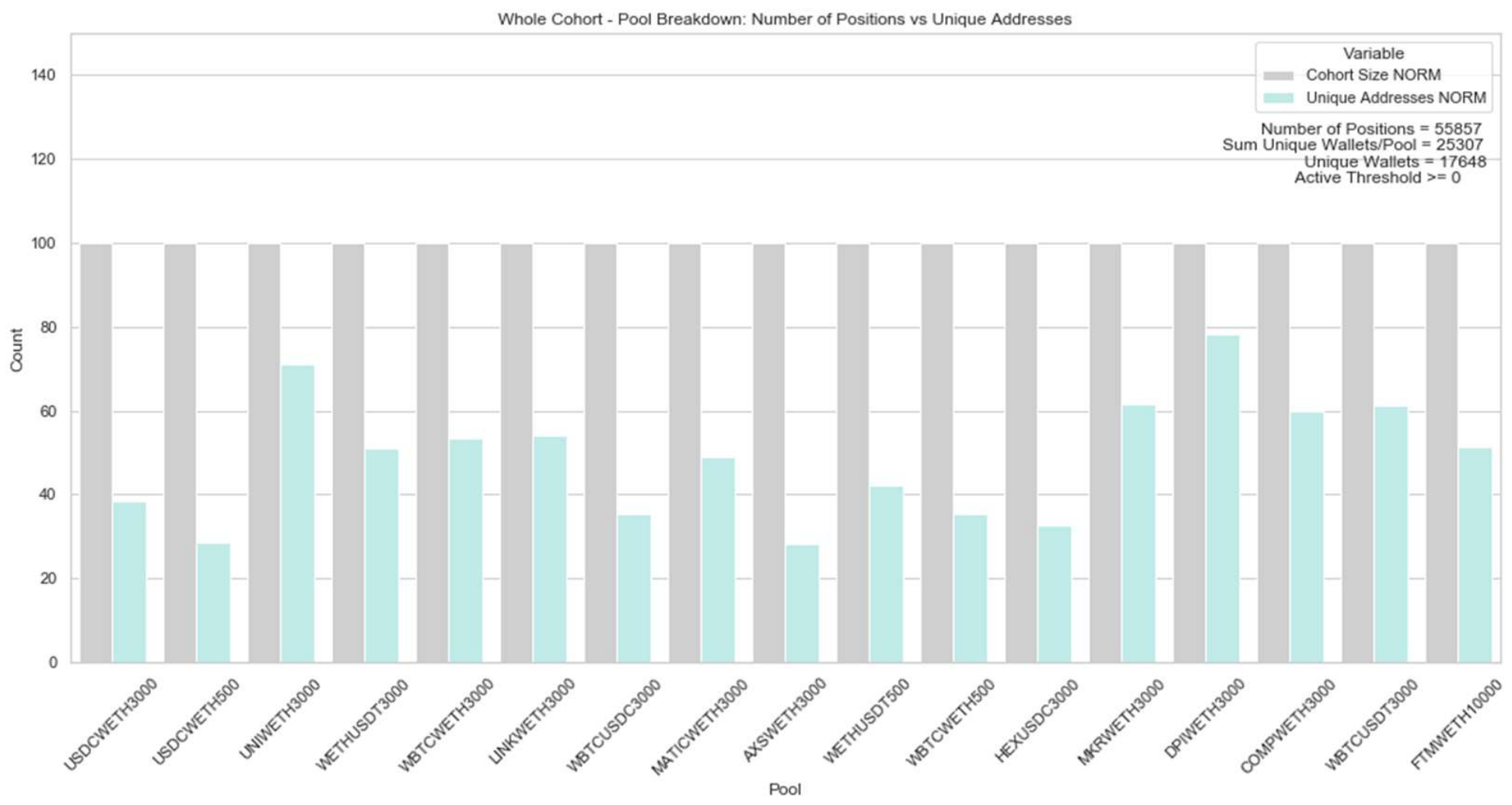

In the chart above we show the number of addresses and positions for specific pools, normalized to the number of positions = 100%. We see that depending on the pool the number of addresses is 20% - 80% of the number of positions, so in average each address runs 1.25-4 positions, depending on the pool.

Pools Analyzed

The charts below show the average and current TVL of the pools that we have already seen in the pie chart above, as well as the fees earned.

Average TVL

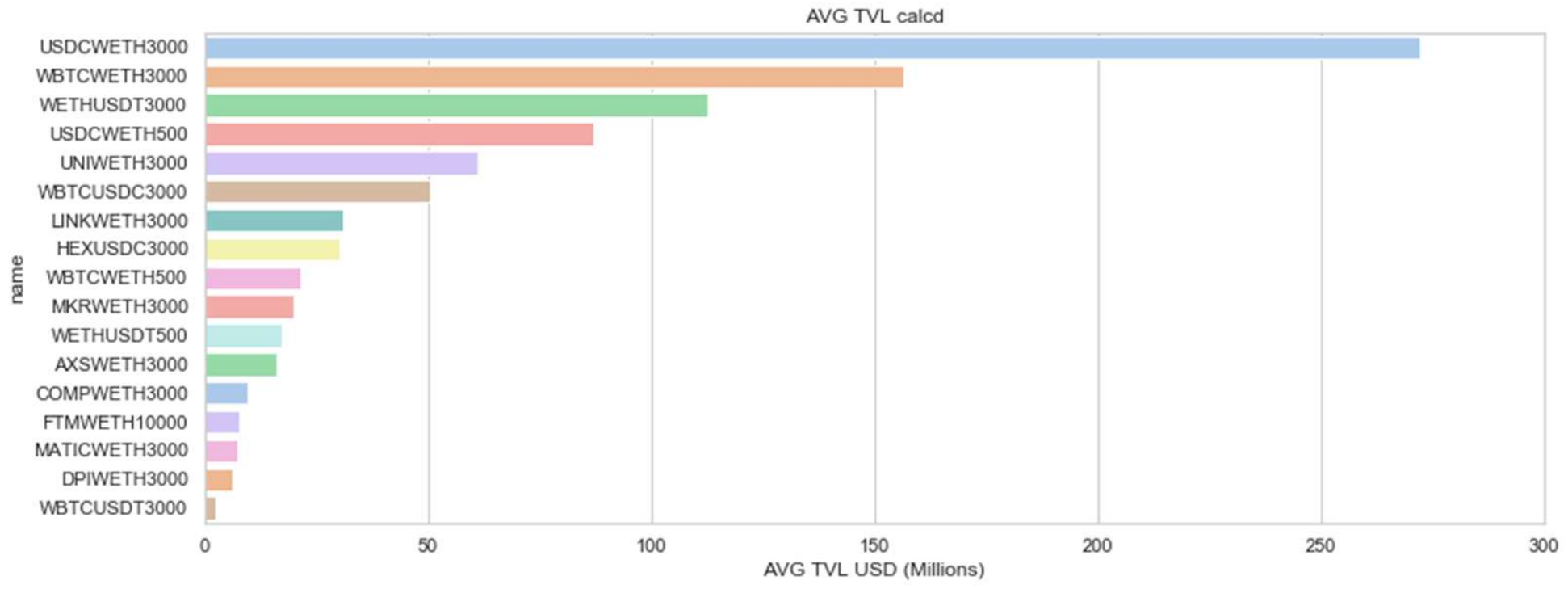

Total Fees

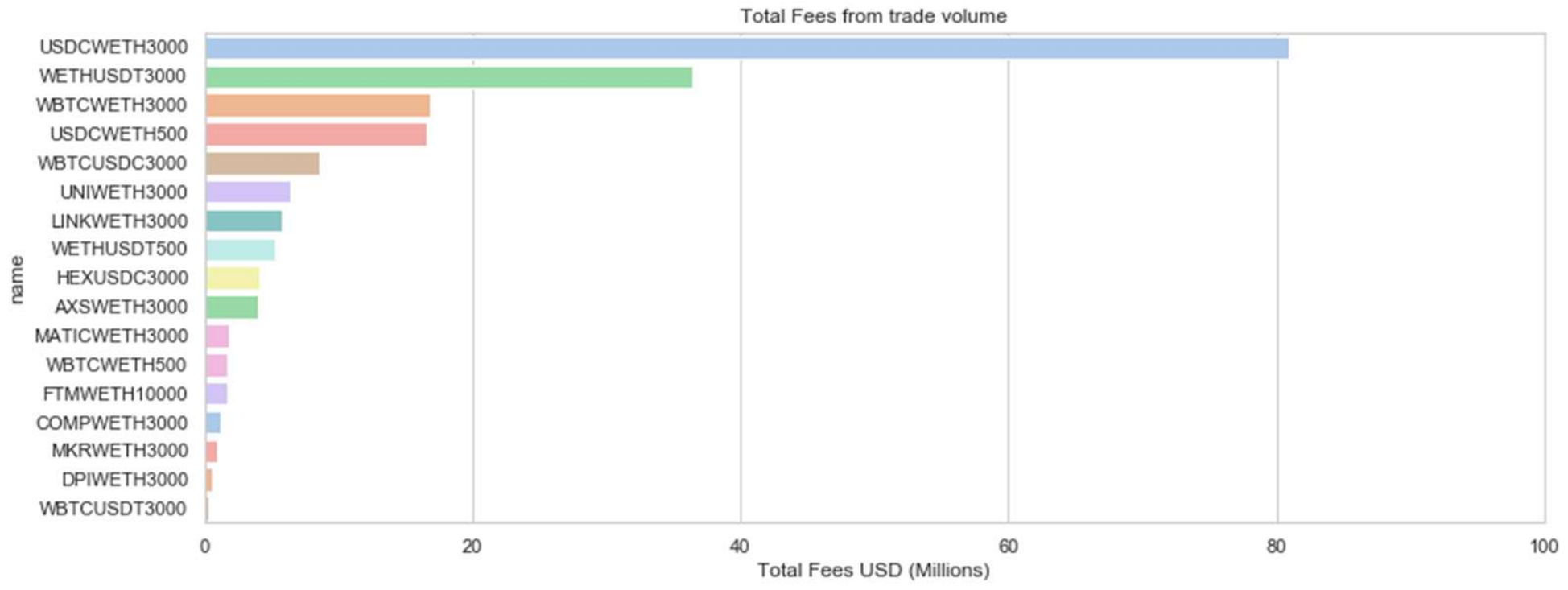

The chart below finally shows the ROI of those different pools. This is a percentage number and is not annualized. The ETH USDT pools start at about 30% ROI, with the 5bp ETH USDT pool still remarkably at over 30% ROI because of its very high trading volumes. After that the ROI numbers decrease quickly, and in the tail the average ROI is under 10%.

Percentage ROI before IL (not annualized)

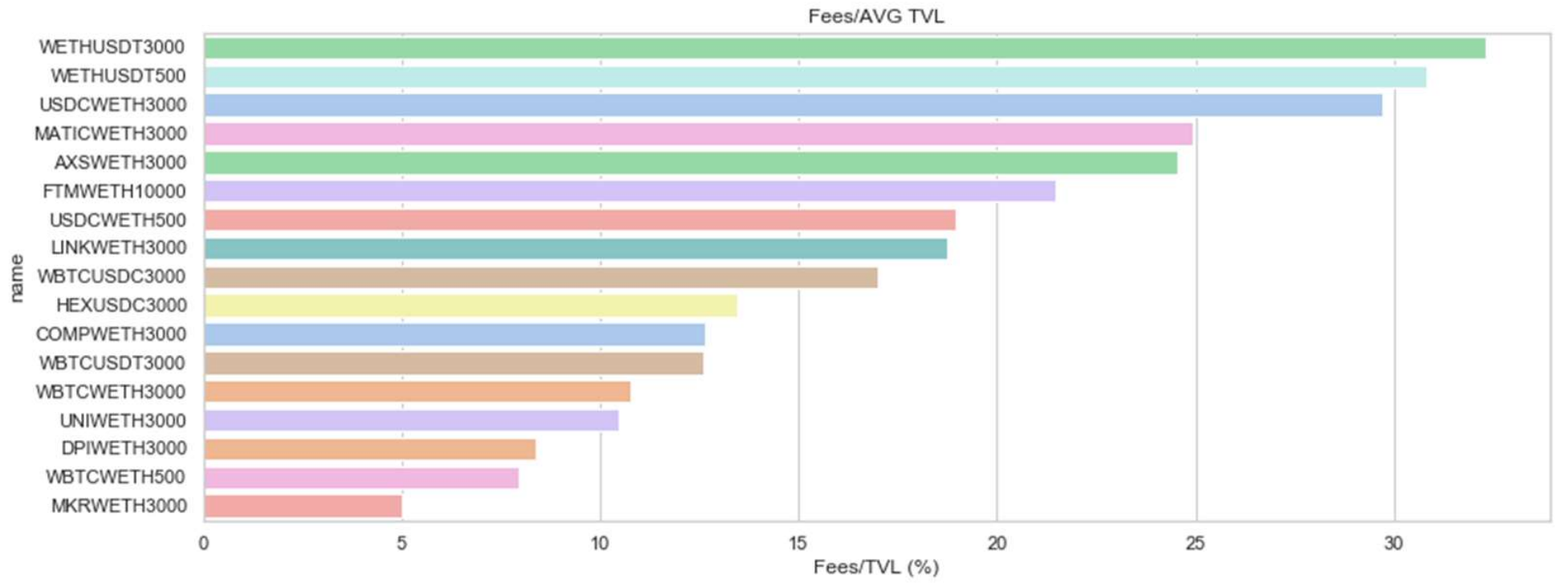

Our key question is now: how much of a dent will IL make into those returns? The short answer is: a lot, to the extent that on a global level, IL wipes out all fees earned by Uniswap v3 liquidity providers. In other words, as a group they would have been better off HODLing than providing liquidity on Uniswap v3. That is what we will discuss in what follows.

## Impermanent Loss Analysis

### Fees vs Impermanent Loss

We have seen in the descriptive statistics section that Uniswap v3 LPs earned a substantial amount of fees both in absolute terms as well as in terms of ROI. However – we know that LPs also suffer from Impermanent Loss (much of which is often permanent). A quick reminder: we define Impermanent Loss as the loss against HODL, and we are using the "novation" methodology discussed above as a means of aggregating IL across pools and time. Also as discussed above, we look at two types of IL in the Uniswap system: the *minimum IL* which is the minimum possible IL a sophisticated investor can achieve by meticulously managing their position and removing liquidity when it goes out of range, and the *actual IL* that Uniswap LPs have actually suffered.

Herein, we calculate the fees from the actual transaction data, where we convert into USD whenever a user decides to withdraw their fees; not-withdrawn fees are converted at the latest available exchange rate. This method is benchmarked and validated against the fees inferred by the published trading volume converted to USD at the transaction time. As discussed above, for the Impermanent Loss we use imputed positions that effectively novate a position every time it is adjusted. This however only gives us percentage IL, and to aggregate it we need to convert it into USD. We perform the conversion at every novation date when an imputed position is closed, and at the last available exchange rate where positions are still open.

Aggregate fees per pool

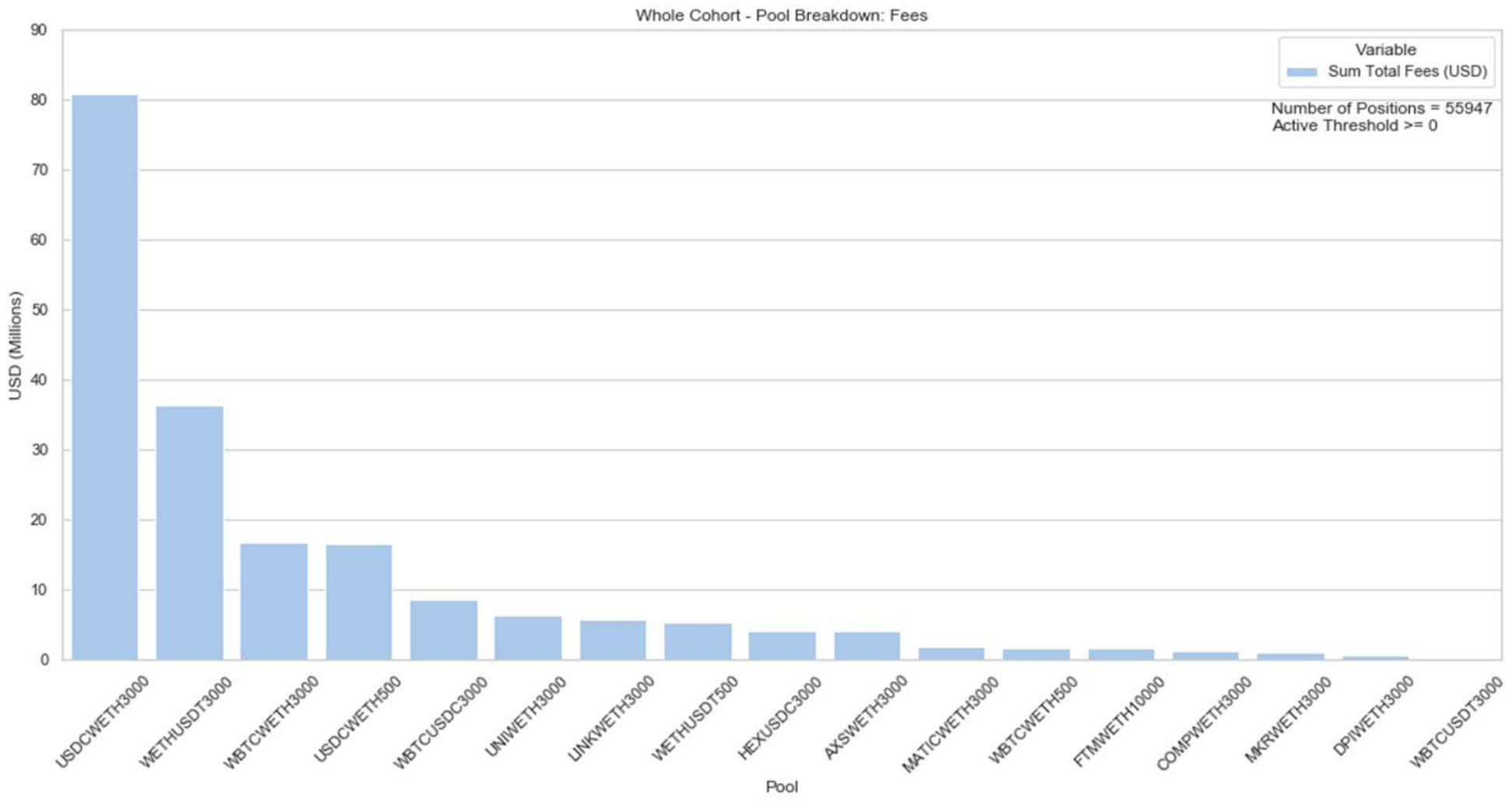

This is to be compared with the actual IL incurred that we plot in the chart below.

Actual Impermanent Loss

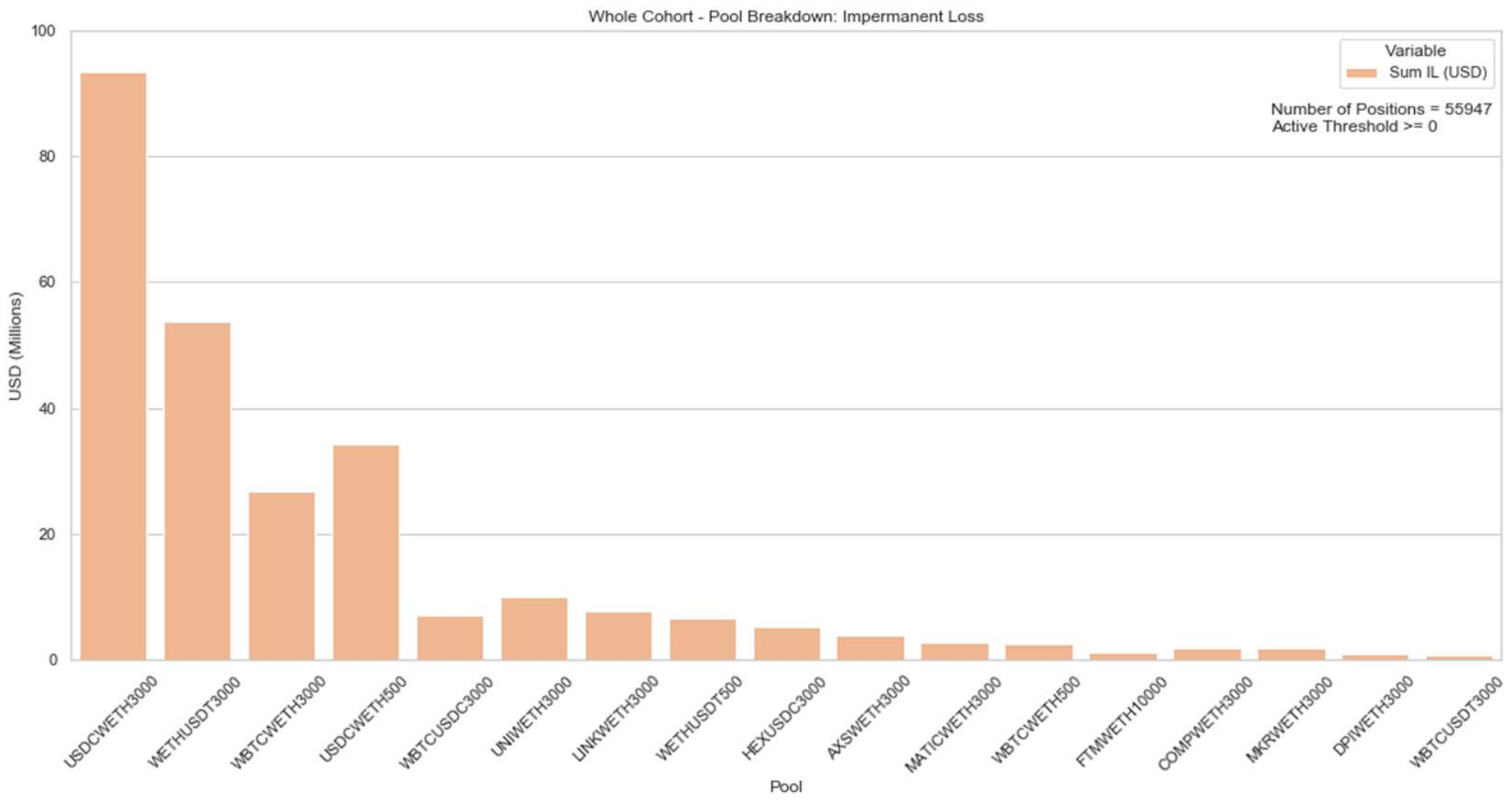

Below we put the two charts together into one. The second of the charts is the same as the first one, except with a different scale to allow for examination of the smaller pools.

Fees vs Actual IL (whole cohort)

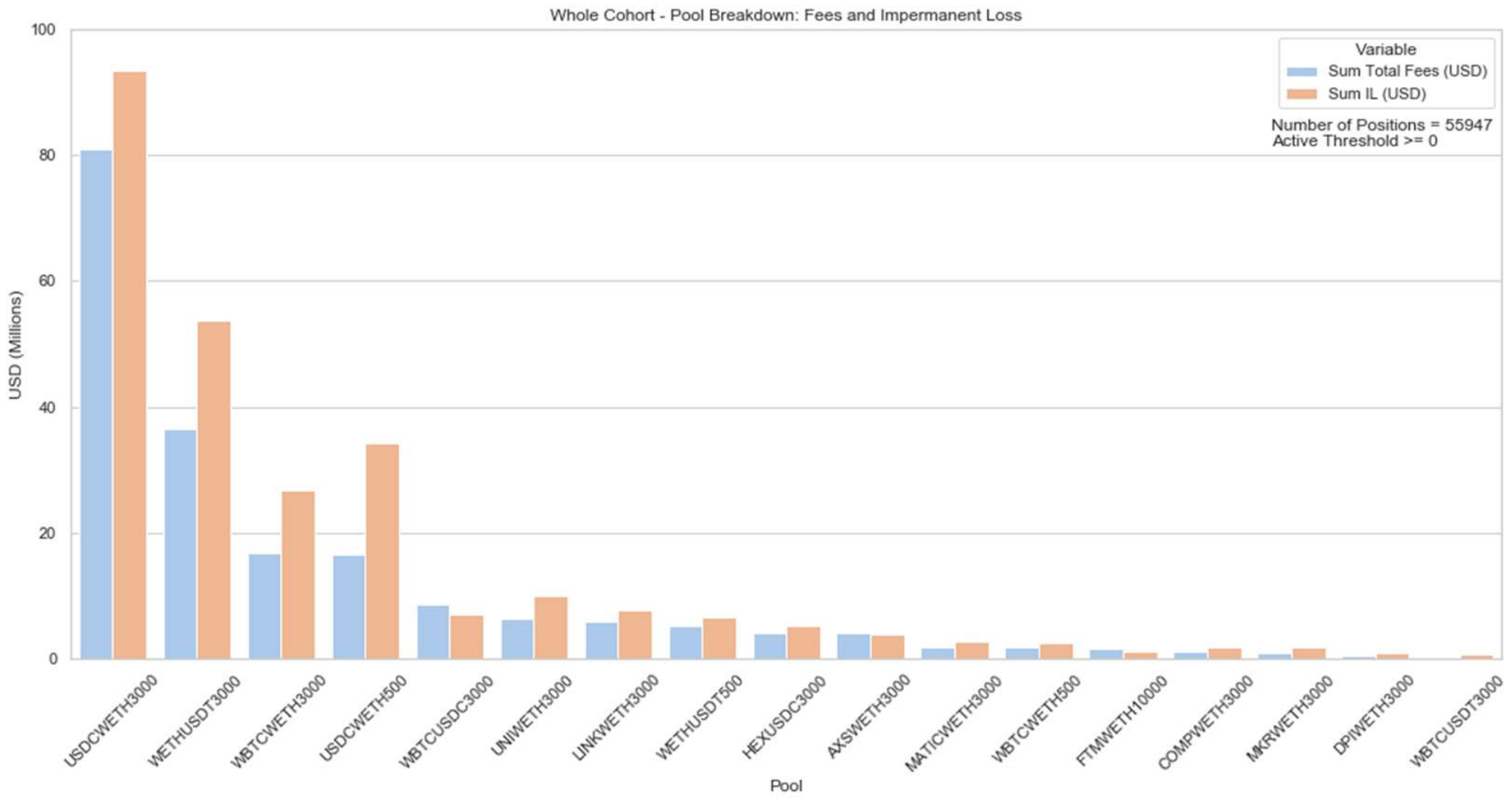

Fees vs Actual IL (whole cohort, rescaled)

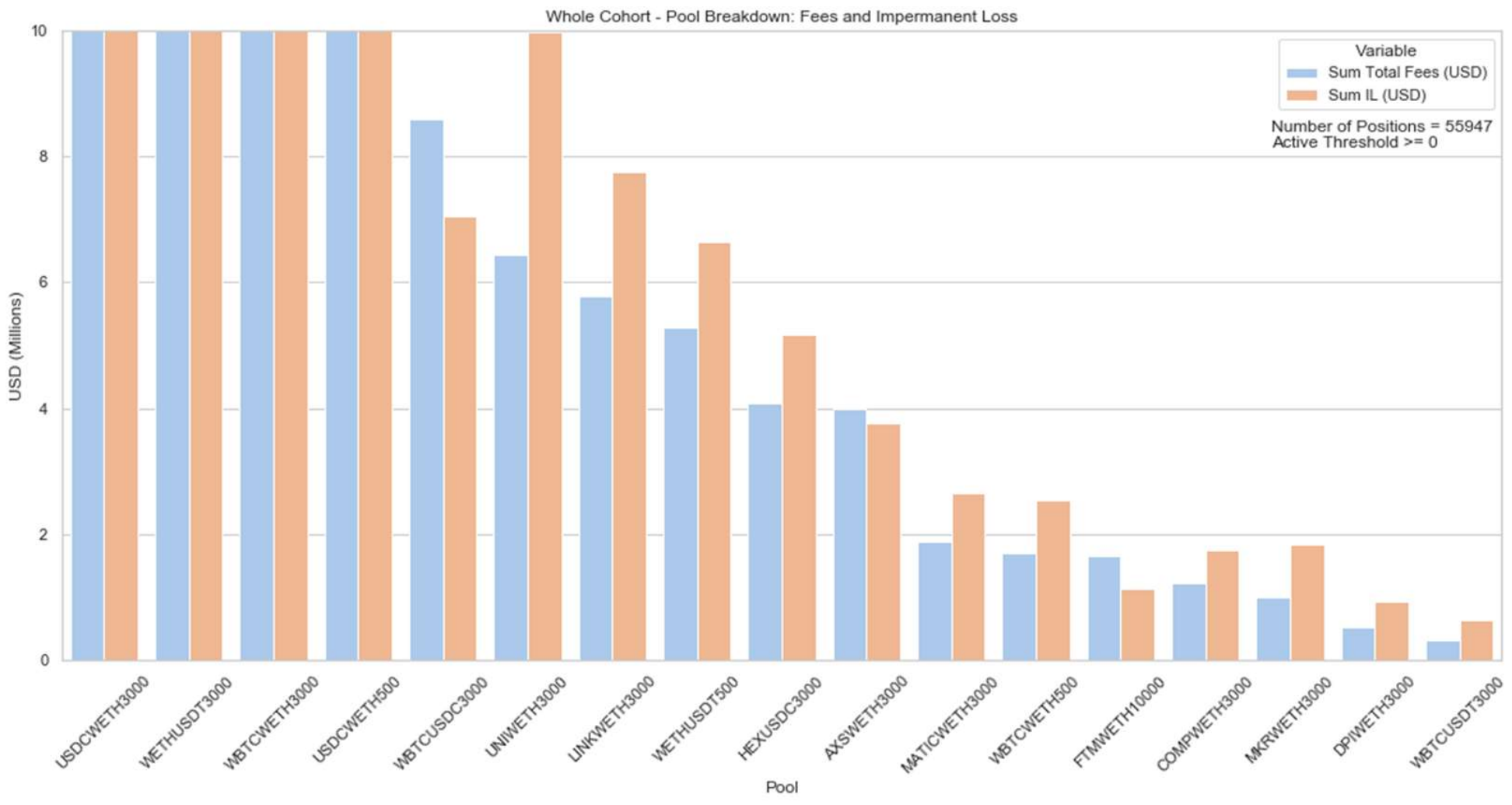

The way to read the above charts is as follows: the left-most pool (USDCWETH3000) earned a little more than $80m since inception. However, during the same time, the IL in the pool is over $90m, meaning that if Uniswap LPs had just HODLd their initial LP positions in the same composition as they contributed it, they would have, in aggregate, made $90m. So, after fees the LPs in Uniswap's biggest pool are still $10m+ worse off than had they just HODLd.

This experience is not unique to the USDCWETH3000 pool. In aggregate over all the pools above, the fees earned were $199m whilst the IL was $260m or 130% of the fees earned – so Uniswap LPs in those pools would have been $61m better off had they HODLd.

Out of all the pools above, only two of them have fees higher than IL – AXSWETH3000 fees are $4m, and IL is $3.75m leaving a modest profit of $0.25m for all LPs together, when compared to the fees accrued. Similarly, FMTWETH10000 where the fees are $1.6m and the IL is $1.1m yielding a profit of $0.5m.

Minimum vs actual IL

Above we discussed minimum vs actual IL. As a reminder – minimum IL is the IL incurred whilst the position is in range, so this is the IL that is unavoidable to incur when wanting to earn fees. The actual IL is the IL that the positions in fact suffered, and that includes the IL that has been incurred outside of the range where no fees have been earned and where a rational LP would have cut the position. In other words – the minimum IL is the minimum IL that an investor who monitors their positions 24/7 would be able to achieve, on positions that are sizeable enough so that the gas costs of the necessary adjustments do not matter. In the charts below the minimum IL is marked by the white line.

We should also point out there is a sweet spot here in terms of adjustment: if the position is too small, the gas costs will be too high to make rebalancing worthwhile. If the position is too big however then rebalancing it back into the initial HODL position may incur substantial price slippage. The reason for this is that the position will, by construction of Uniswap v3, be 100% in the recently underperforming asset that has just breached the lower bounds of that range. *Rebalancing* means selling this asset against the asset against which it just has underperformed. In other words: rebalancing may mean to sell into a falling market, the additional costs of which can be substantial.

Fees vs Actual and Minimal IL (whole cohort)

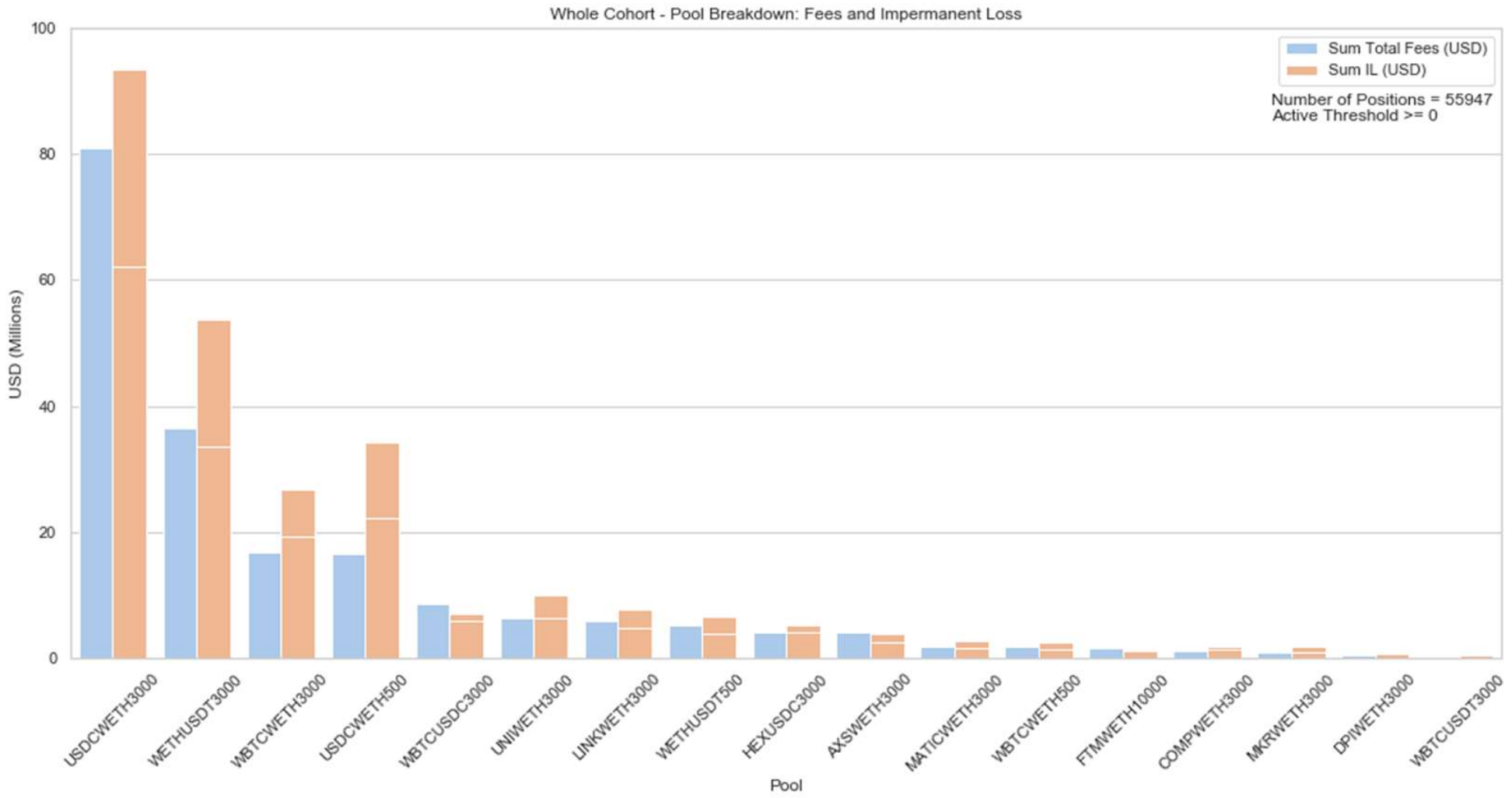

Fees vs Actual and Minimal IL (whole cohort, rescaled)

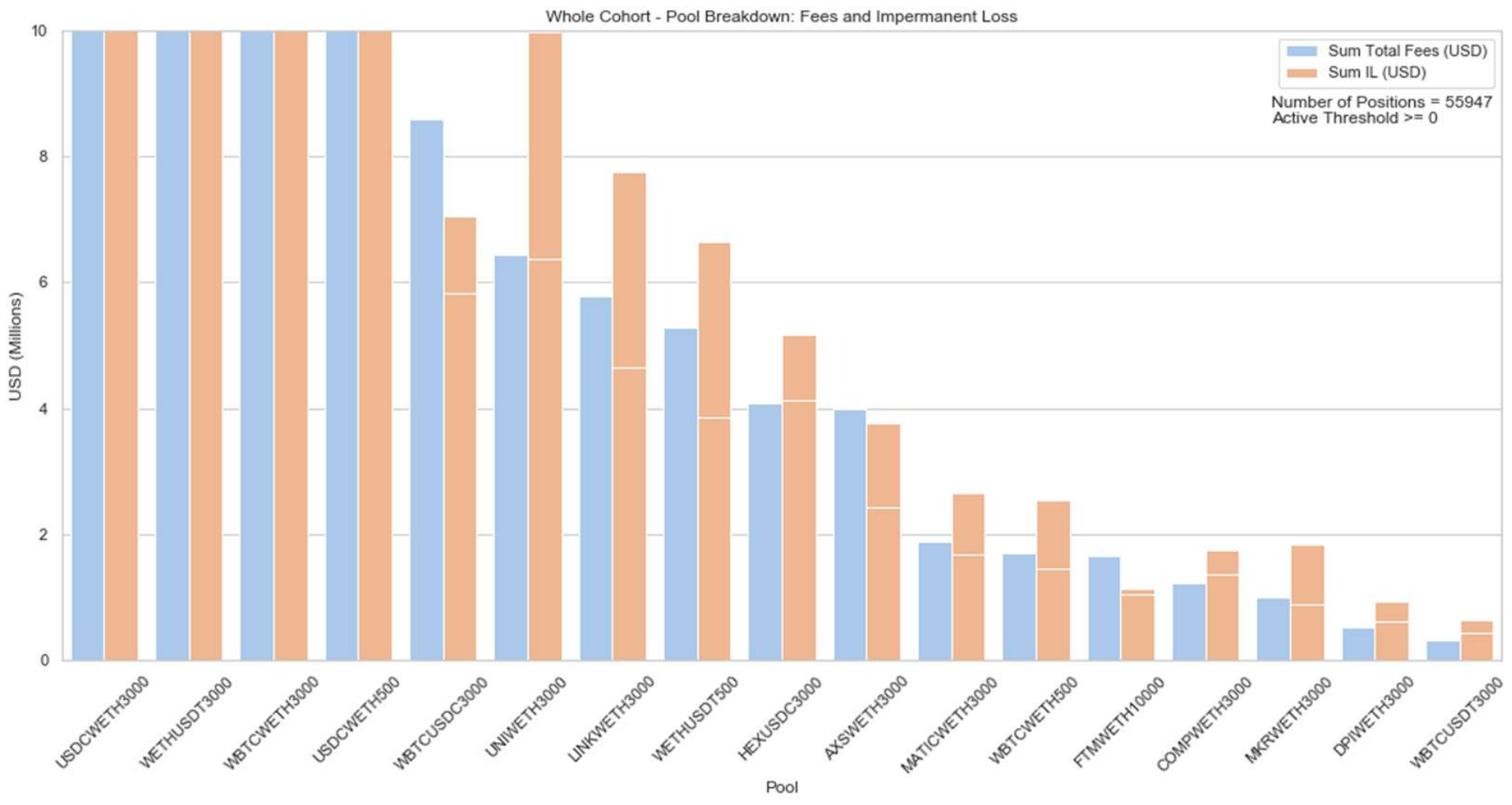

Another thing we would like to allude to here – but which is the topic of forthcoming research – is that rebalancing positions locks in IL, meaning at this stage it is definitely no longer impermanent. This is very similar to delta hedging an option: whenever the spot moves the trader has to decide whether to rebalance the delta – which incurs transaction costs and locks in the Gamma bleed – or hope that the spot returns to its initial value at which stage the position is again perfectly hedged.

The situation is similar here: whenever the spot price leaves the range, a certain IL has been incurred by the LP who then has two options:

1. Do nothing, sit outside the range, earn no fees, and hope the spot returns to its initial value.
2. Cut the position, crystallize the IL (typically 3-10%) and either leave the game, or place at the current range to have another throw of the dice.

We are working on a Monte Carlo simulation and some theoretical analysis to better understand the effects of this phenomenon, and this will be subject of forthcoming research.

Segmentation by duration of the position

After finding that on average Uniswap v3 users do not perform better than HODL we then tried to understand whether some groups consistently do better than others, further increasing the losses of the underperforming groups.

Our first analysis was to look at what we called *active* and passive users, where the latter were defined as those who contributed liquidity exactly once, and who either still held that position without any modifications, or who had closed it out in a single withdrawal transaction. When we ran this analysis, we found no statistically meaningful differences between those two groups. This meant that either sophisticated users of Uniswap v3 did not manage to capture profits from their less sophisticated peers, or we did not have the right segmentation to analyze this hypothesis. We realized that our passive group consisted of two very different segments:

- Inactive users who “fire and forget” their LP positions until they eventually withdraw, not necessarily at the optimal point in time, and
- Highly sophisticated users who, for reasons of privacy or others, quickly add and withdraw their positions frequently but do not reuse LP tokens, and maybe not even addresses.

So, in the second state of this analysis, we also looked at the tenure or lifetime of a position, on the assumption that, on average, those who go into the market and leave it shortly thereafter are more sophisticated, or at least more active, than those who stay longer.

Cumulative percentage of active positions vs position lifetime

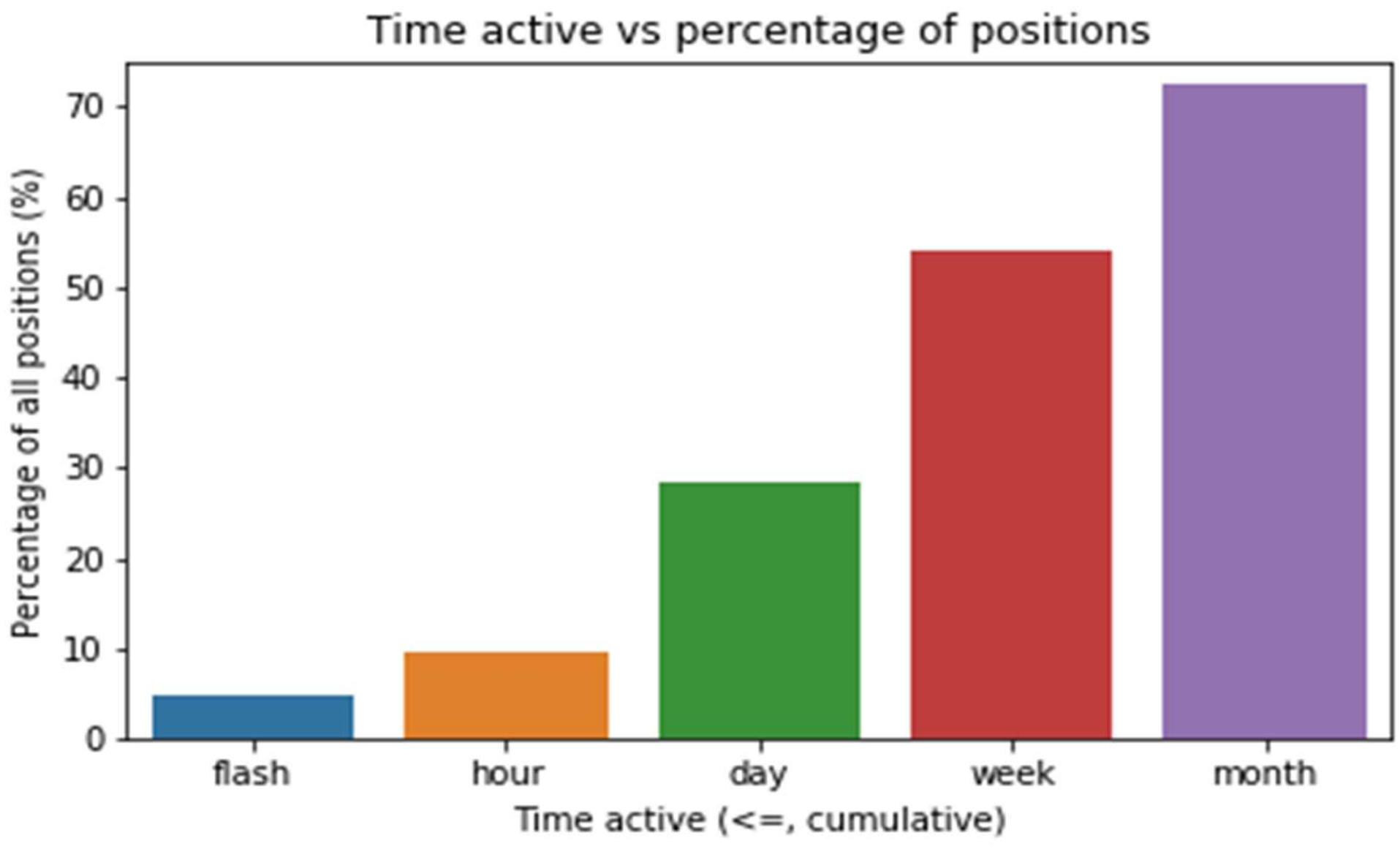

To start with an overview, the above chart shows the cumulative number of positions as a function of their duration. The way to read this chart is that about 5% of all positions are flash positions (they only exist for a single block), another about 5% exists for at most an hour. 20% of positions exist more than an hour but less than a day, another 25% more than a day but less than a week, another 20% in the month, and finally another 30% are longer than a month.

Cumulative percentage of fees captured vs position lifetime

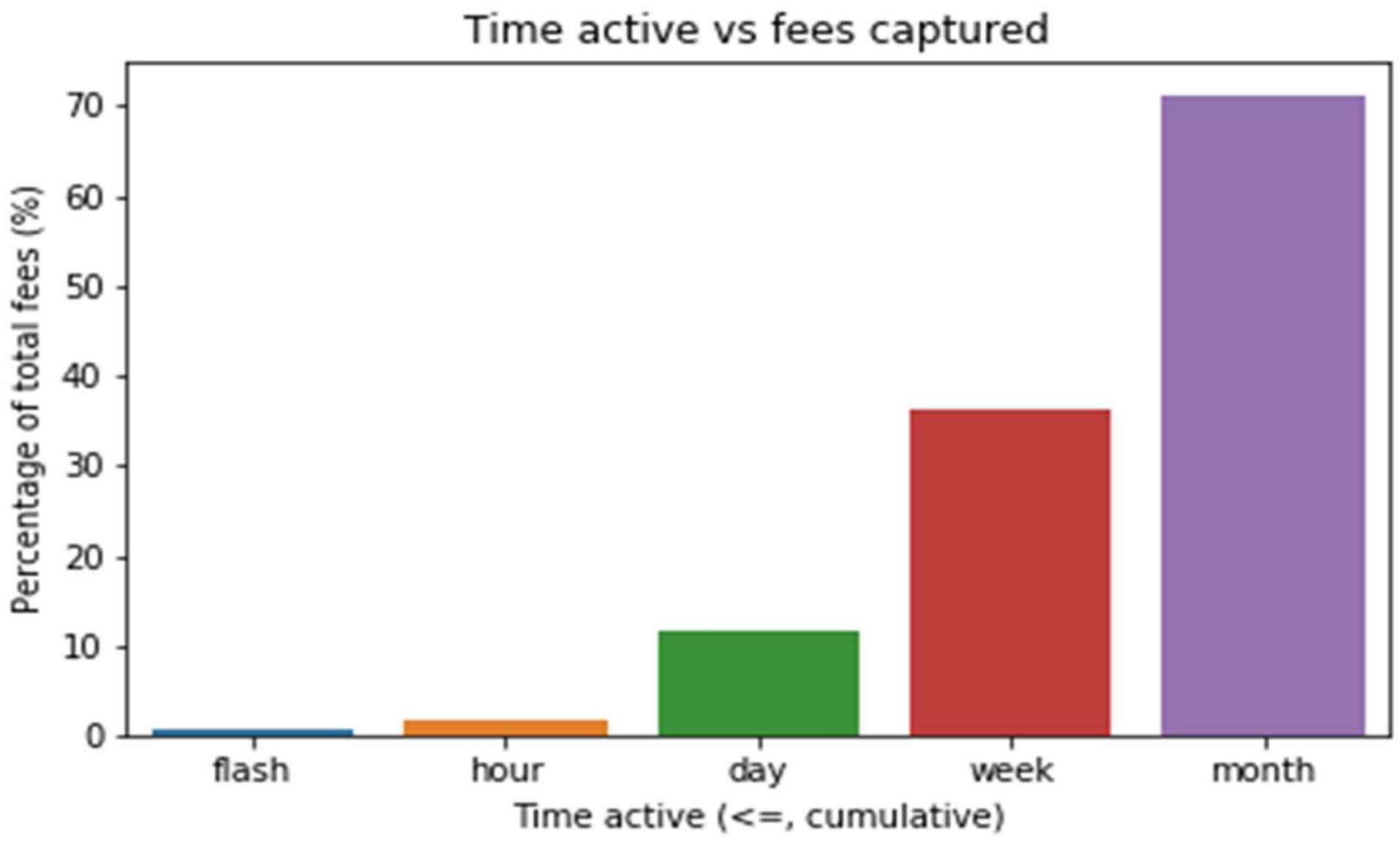

The above chart is the same as the one before it, expect that we are looking at percentage of fees instead of percentage of positions. What is interesting is that the smaller numbers are not as small as one may expect them to be. For example, the “within one week” segment went from ca 55% of total positions to ca 40% of total fees. Given that “within one month” group remained constant, one would expect a division of the total fee share in the "within one week" segment by 4 (number of weeks in a month, rounded to the nearest integer) to about 15% of the total fees rather than the 40%.

On first sight we may expected this to translate into a better risk-adjusted performance of short-term LPs when compared to long-term LPs. However, when looking at the chart below this does not seem to be the case:

Fees vs IL by time active (whole cohort)

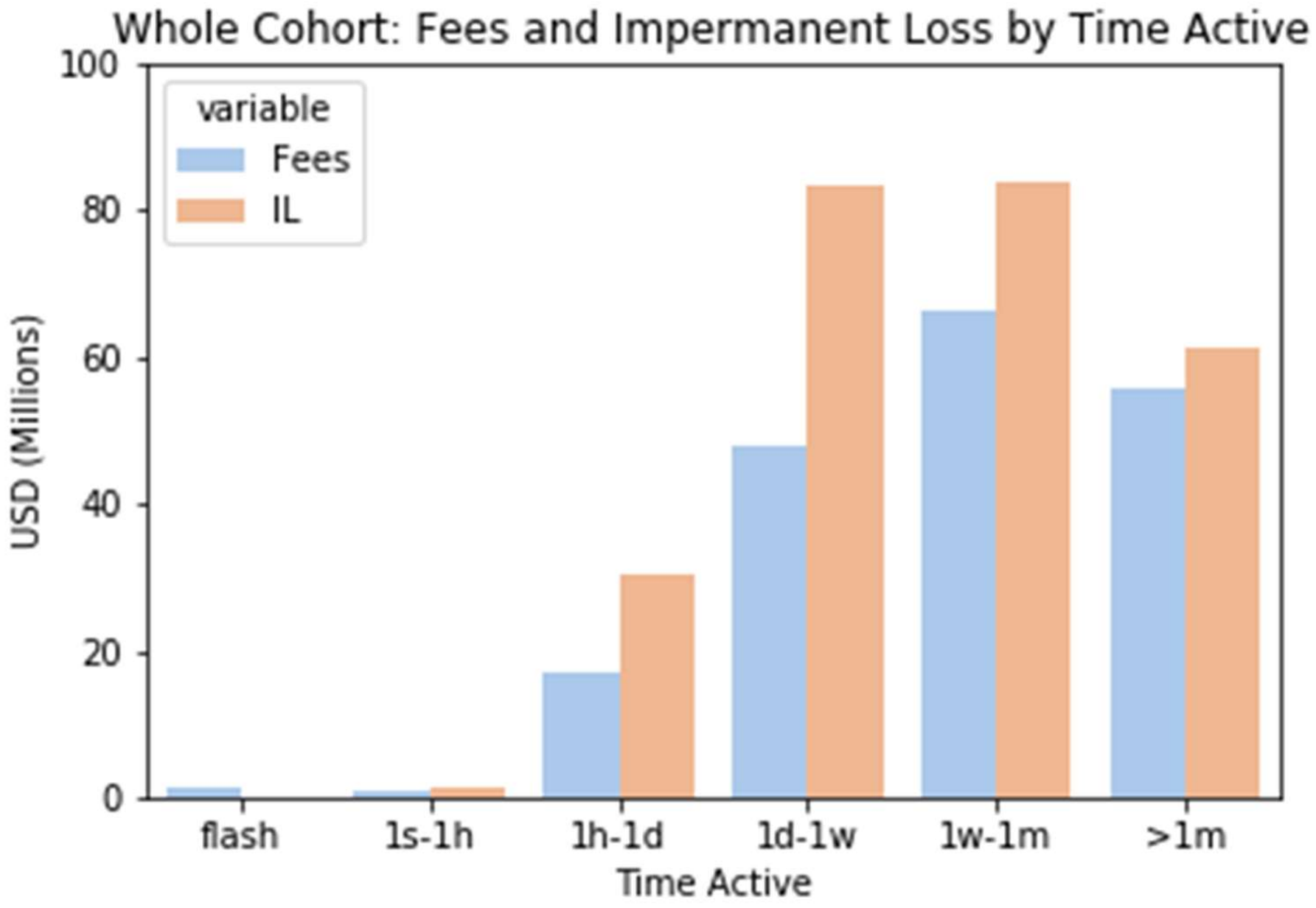

In the chart above we see the fees and IL for each of the segments and we see that out of the major fee earning segments, the *>1m* segment does perform best – but only in relative terms as even for them the IL outpaces the fees.

The only group that we could identify that consistently made money when compared to simply HODLing was flash LPs who only provided liquidity during one block. This behavior has been coined Just-In-Time (JIT) liquidity. The liquidity was provided intra-block and it did not cause any meaningful IL. All other segments have an IL/fees ratio that is greater than 1.

Ratio IL vs fees (whole cohort)

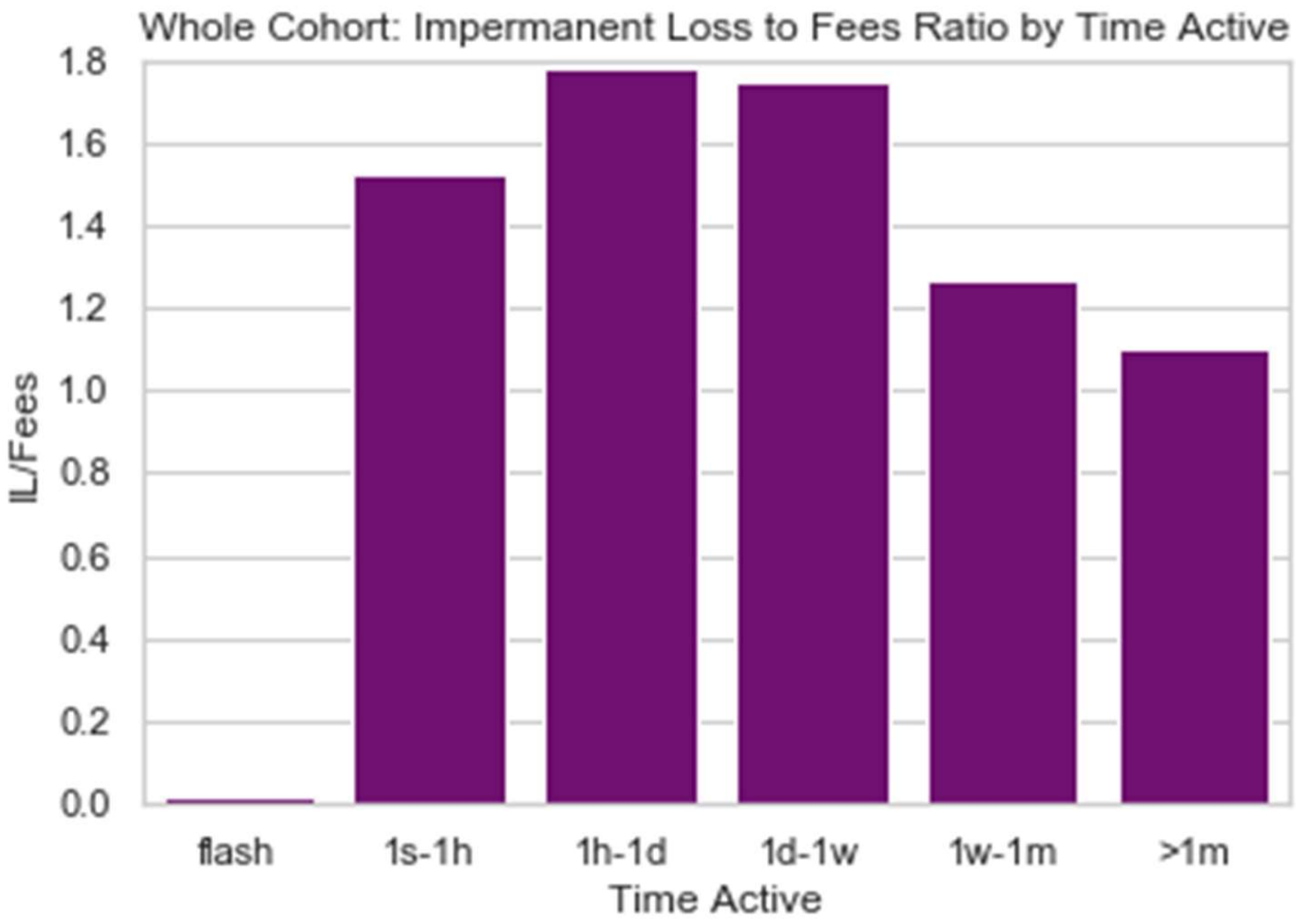

The chart above illustrates this: it plots IL / fees for the various time slices. IL is in a way the cost of running a position, therefore the above chart plots something like the cost / income ratio popular in analyzing traditional financial institutions. Cost / income ratios > 1 mean the institution loses money, and here this is similar: in the 1h-1d slice the ratio went close to 1.8, meaning they lost (compared to HODL) $180 for every $100 in fees, leaving a net loss (against HODL) of $80 for every $100 in fees.

Risk adjusted returns (fees/IL – 1)

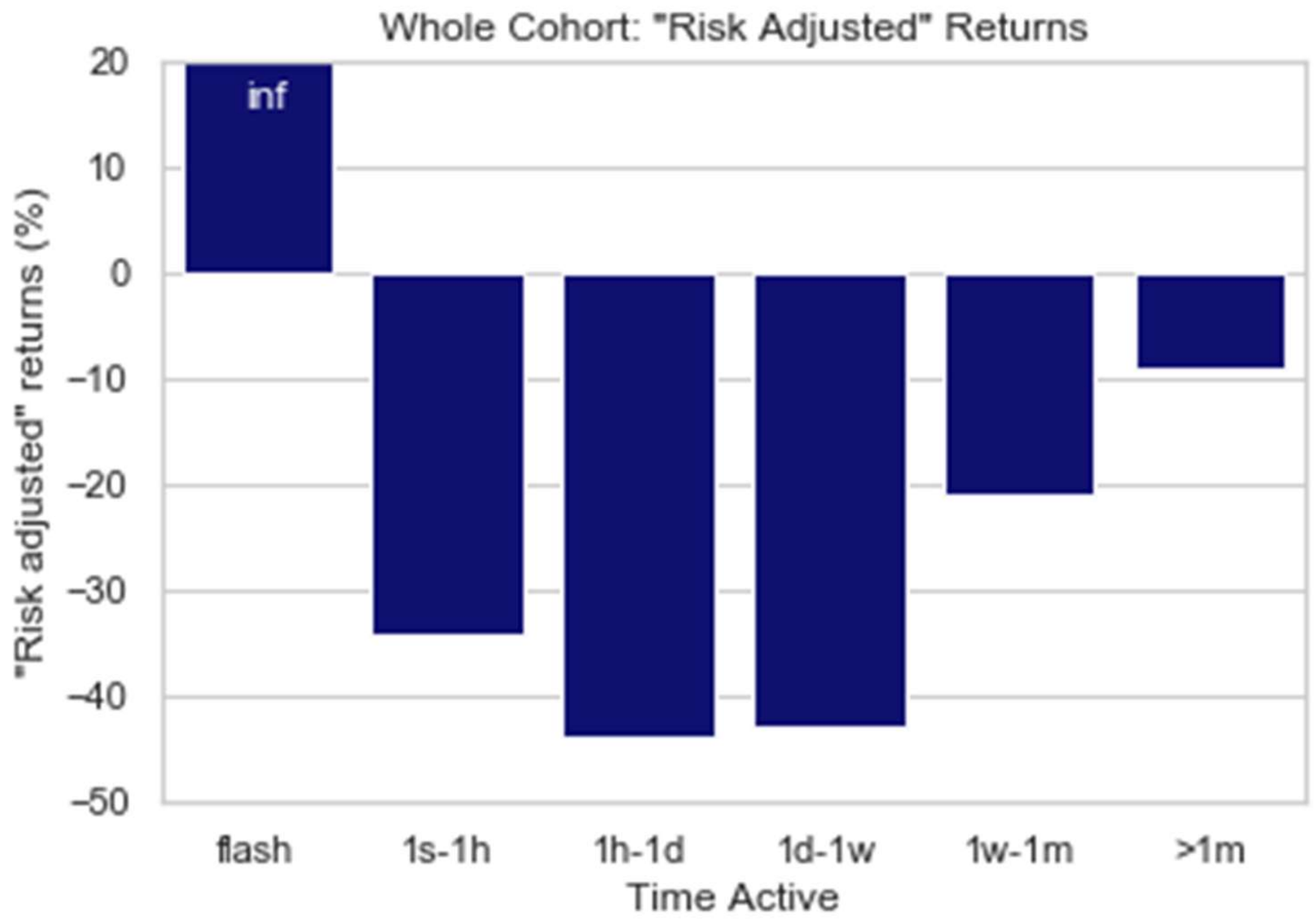

The above chart is a slight reshuffle of the previous one for those who prefer return-style numbers to cost/income ratios. Here we calculate fees / IL (note that the previous chart was the inverse) and then subtract 1. If we consider IL a risk measure rather than a cost, then this is a risk-adjusted return number, or a RORAC (return on risk-adjusted capital). A number > 0 means the positions made money (against HODL), and a number < 0 means they've lost. We see above that all positions lost, except the flash slice where IL = 0 means this measure diverges towards positive infinity.

## Positive vs. Negative Returns

In this section we are aggregating the positions by wallets and are segmenting them by whether the overall returns of the wallet were positive (green), or whether they lost money (red). Positions with less than less than $1 of liquidity were removed from this analysis.

Percentage of wallets with pos/neg returns

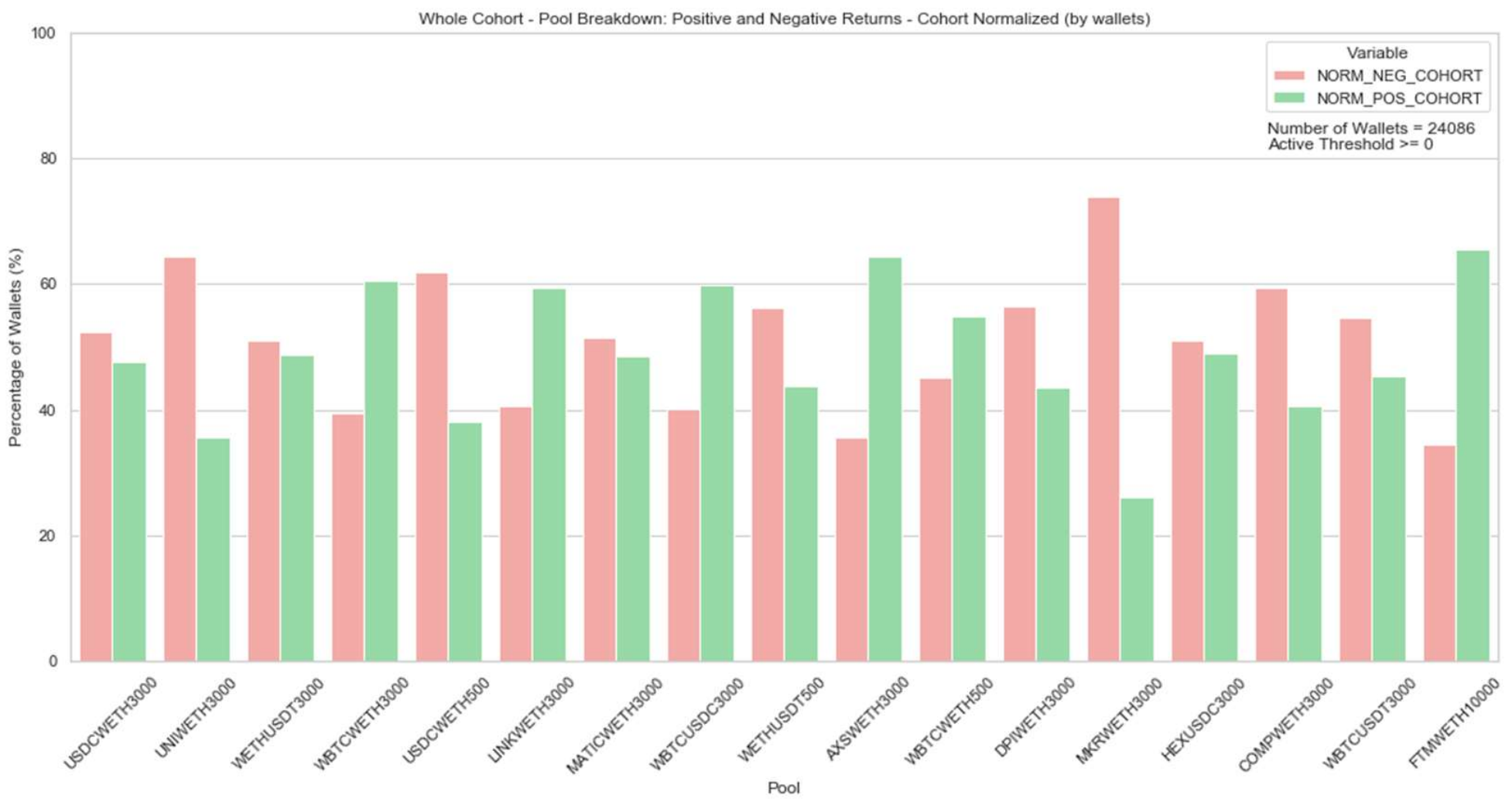

Above we show the percentage of wallets per pool with positive vs. negative returns. Looking across the different pools, the percentage of users incurring negative returns is generally between 40-60% per pool. On the low end, in the FTM/ETH pool, roughly 34% of users have negative returns, while on the high end, in the MKR/ETH pool, roughly 74% of users have negative returns.

Mean USD returns per wallet in positive and negative segments

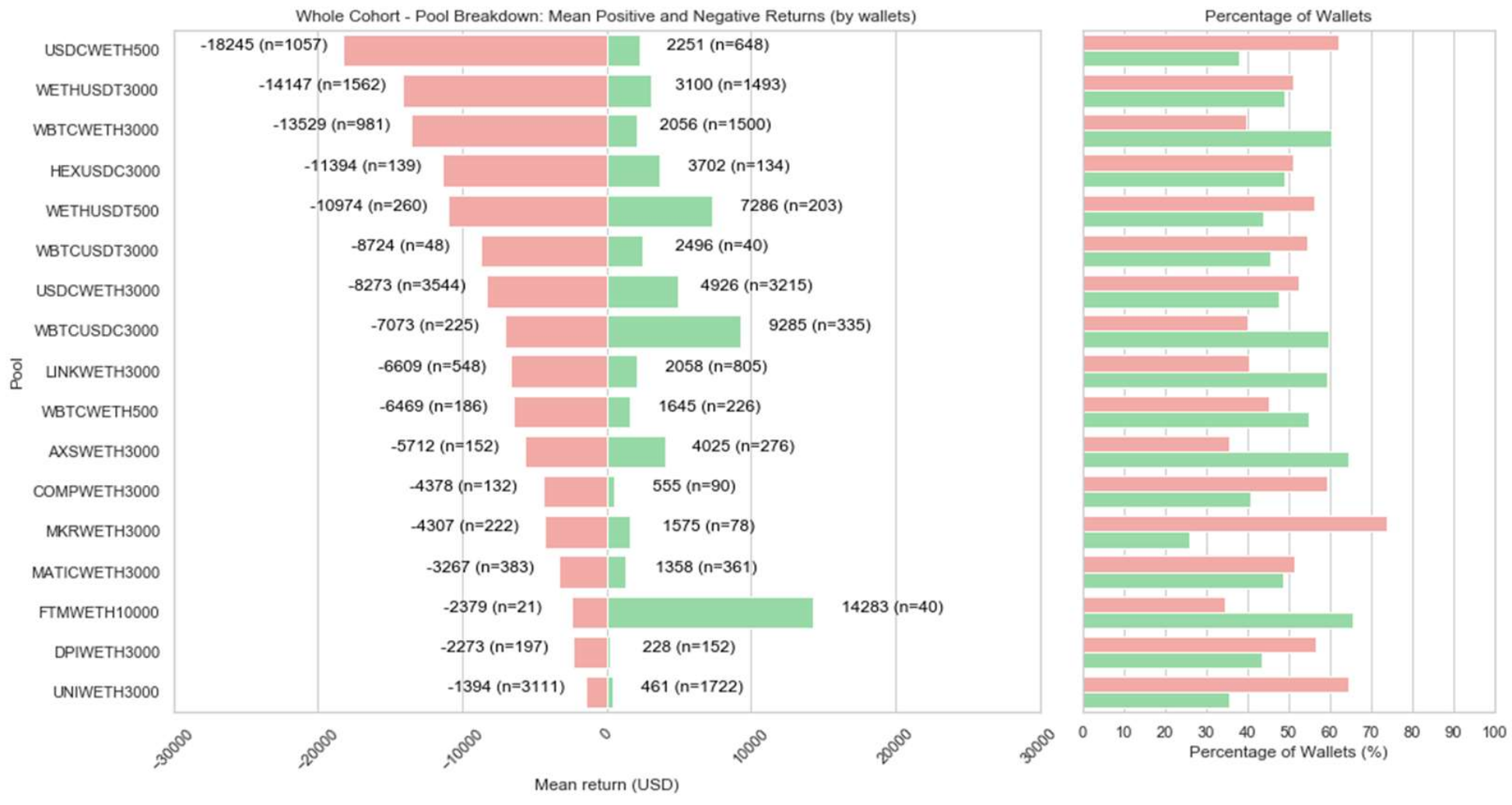

The above chart is a bit more complex so we describe it in pieces. The key information is on the left side of the chart where we look at the mean return (in USD) by wallet and pool, separated by wallets that lost money (red) and wallets that made money (green). The number of wallets ("n") in each group is also shown. The chart on the right side gives the percentage of wallets that made or lost money per pool. For example, we can see in the USDCWETH500 pool that 1,057 LP wallets or 62% of LPs in the pool have lost money, with a mean return of -$18,245, whereas 648 LP wallets or 38% of LPs in the USDCWETH500 pool earned money, with a mean return of $2,251.

The first thing to take away is that the red bars tend to be longer than the green bars, so the average loss is bigger than the average gain which is consistent with the average LP losing money. The WBTCUSDC3000 and FTMWETH10000 pools are outliers here, where not only the average gain was higher than the average loss but also where the number of wallets making money is higher than those losing money. That is of course consistent with these pools being some of the few pools that make money for the average LP. It should also be noted that these pools are relatively small (1-6% TVL) when compared with the other pools analyzed.

Here we look at the percentage of wallets with positive vs. negative returns by the amount of capital contributed. Within each group, the percentage of wallets with negative returns is larger, and the percentage of wallets with negative returns grows with the size of the capital contributed, suggesting that larger LPs are suffering higher IL. We attribute this to larger LPs spreading their positions across wider ranges and therefore collecting fewer fees.

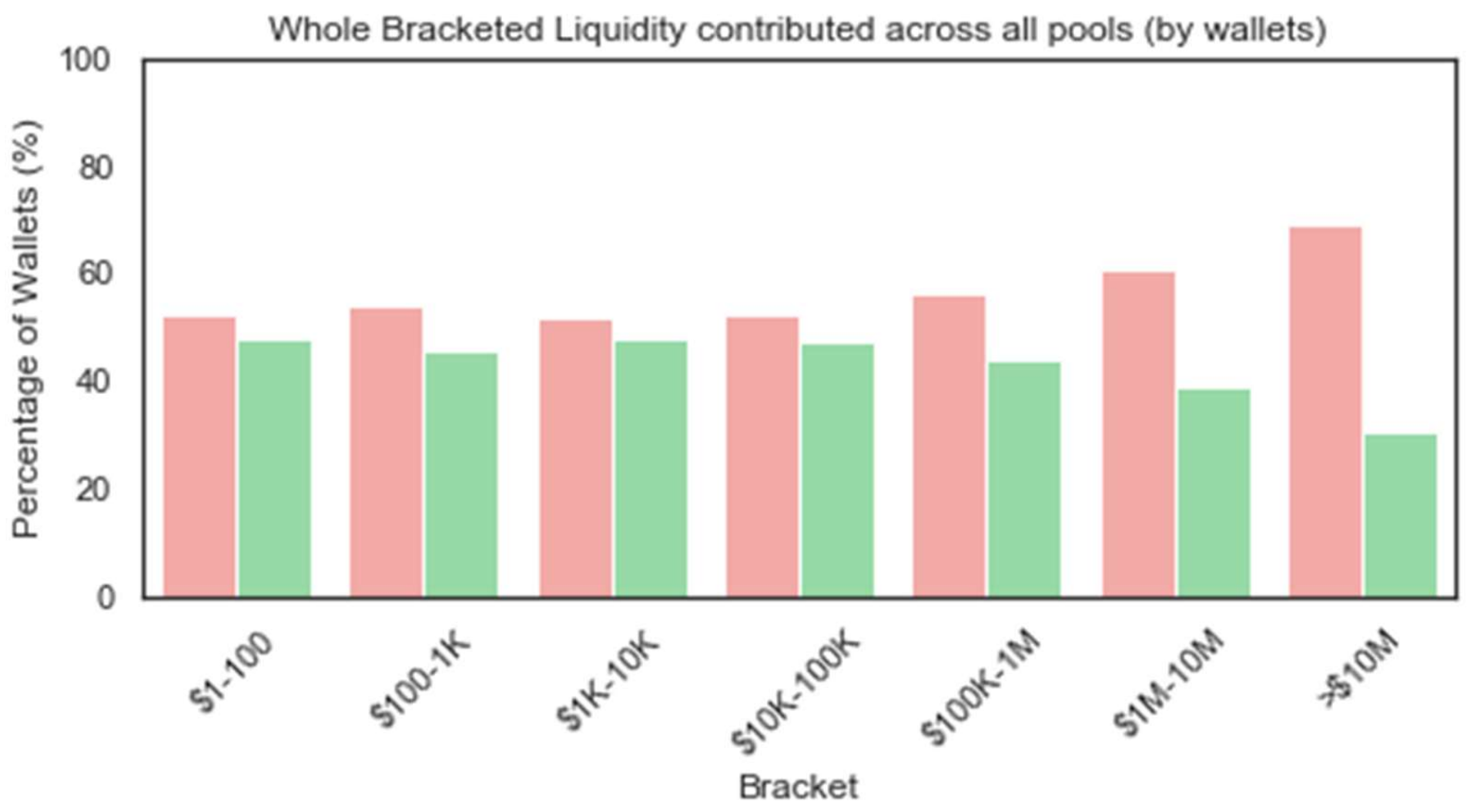

ROIs

In this section we are looking at average ROIs.

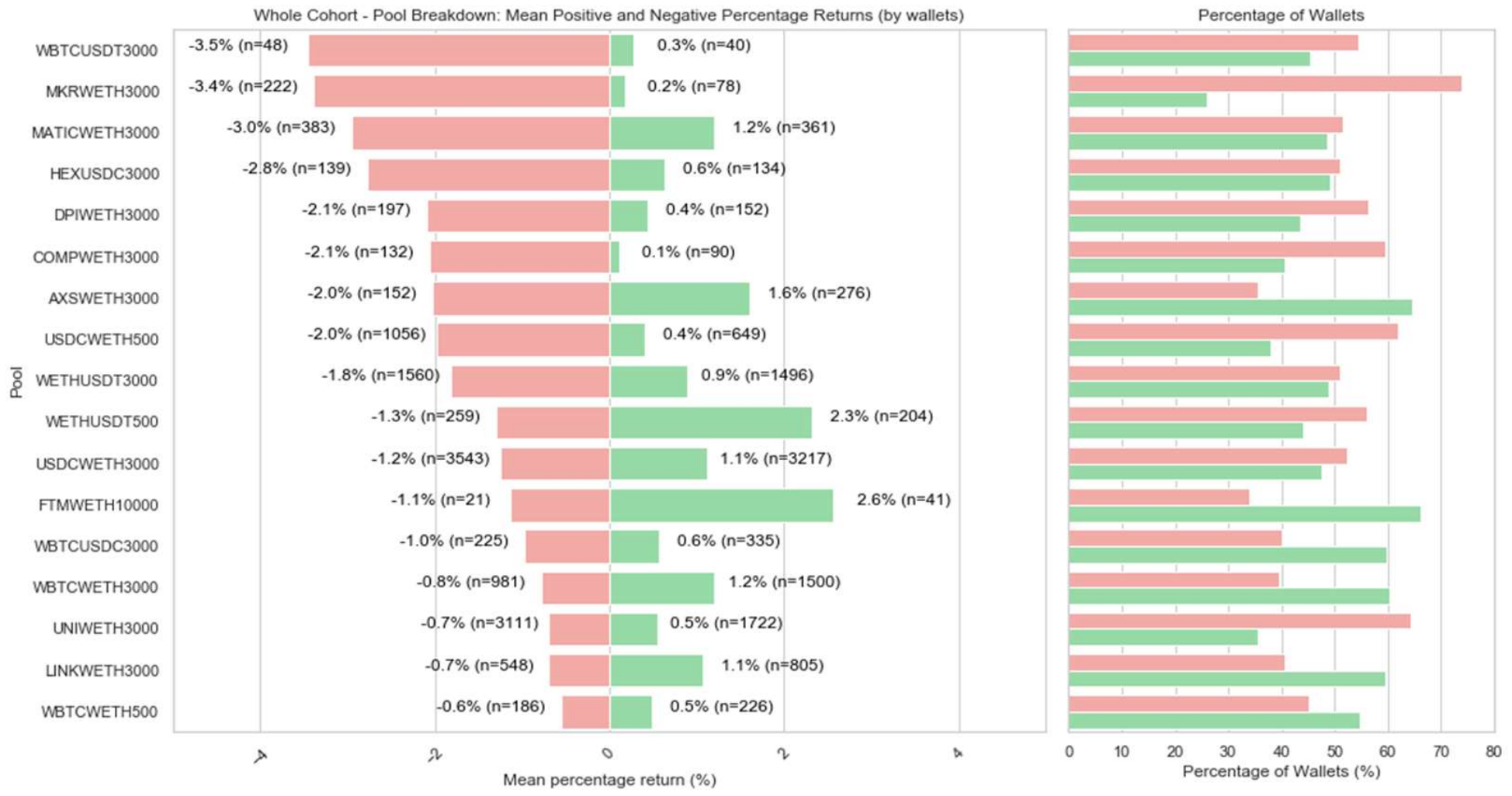

We have seen a similar chart above where we looked at the average USD returns. Here we are instead looking at the average ROIs [(Fees - IL) / time-weighted average liquidity]. These ROIs are not annualized. We again see that the red bars – the average ROI for the segment of users with negative returns – tend to be bigger than the green bars, which represent the average ROI for the segment of users with positive returns. Also, the numbers are generally relatively small: on the positive returns side, the best average return per pool is around 2.6%, for FTMWETH10000. That is a nice return for a quarter, but not overwhelming in the grand scheme of things. We then have five pools who return an average above 1%, and the remainder of pools are below 1%, all the way down to 0.1-0.2%. In the red segment, the median is about -1.8%, and the range is from -0.6% to -3.5%.

## Conclusion

In this paper we have looked at the economic performance of the leveraged AMM design pattern, as implemented by Uniswap v3. The economic performance of every AMM has three components:

1. Fee income
2. Gas cost
3. Impermanent loss

The first component, fee income, has been well studied as it is commonly available on the Uniswap UI and aggregated in various public dashboards. We found that, while gas costs appear high, in practice they have little effect on the outcome of the cases we have studied. It is likely that this is a case of self-selection. Gas costs are a highly transparent and upfront "cash" cost, so potential LPs self-select and will simply not engage in a transaction should the gas costs be prohibitively high. However, it was found that Impermanent Loss (or "IL") is the dominant factor in determining the financial impact of liquidity provision on Uniswap v3. Importantly, the primary revenue generation - the fee income – is all but entirely eclipsed in aggregate by the IL; all fees earned by Uniswap v3 participants are rendered inconsequential, on average, in the face of the IL to which they are exposed. To the best of our knowledge, this publication is the first to explore these data in detail.

The findings reported here go against the widely held belief that leveraged liquidity provision offers a reliable means to mitigate IL. We have made a deliberate attempt to identify sub-groups that consistently outperform others on Uniswap v3 and have been unsuccessful. Therefore, we were unable to confirm our hypothesis that "*active*" LPs (those who are dynamically managing their position) are more successful than "*inactive*" LPs (roughly defined as those who interact with their position infrequently, regardless of whether their liquidity is in range). Within a reasonable margin of error, the performance of both groups is indistinguishable. It could be that our inactive cohort is in fact an amalgamation of two distinct types of players: those that withdraw their position quickly and open a new one (possibly from another address) and those who stay in their static positions for a long time. By that same token, we cannot be certain that all interactions with a liquidity position are motivated by a desire to optimize its performance.

In short, the user who decides to not provide liquidity can expect to grow the value of their portfolio at a faster rate than one who is actively managing a liquidity position on Uniswap v3. In fact, only 3 of the 17 pools analyzed earned fees that exceeded the IL, and only by a small margin. Those 3 pools were mid-sized pools of about $2m, $4m, and $8m fee income, and the outperformance was de minimis ($2m in aggregate profits with aggregate fees of about $14m, and aggregate IL of about $12m). It must be said that the newly born industry of liquidity optimizers are facing an apparently insurmountable problem. Certainly, our findings suggest that a winning strategy could be formulated (although it seems that no one has yet), and it is reasonable to assume that one among the plethora of new projects claiming to have the proverbial secret sauce may actually achieve it. However, the expected returns may be comparable to the annual rates offered by mainstream consumer banking products.

The time that a position has remained open, and its correlation with fee revenue as a proportion of IL is worth further comment. The only group that we could identify that consistently made money when compared to simply HODLing was flash LPs who only provided liquidity during one block. This behavior has since been coined Just-In-Time (JIT) liquidity. In this group there was no meaningful IL. In the other groups the IL / fees ratio was always >100%, meaning they lost out. In fact, it went as high as almost 180%, meaning a loss of $180 for every $100 earned in fees, leaving a net loss of $80.

JIT liquidity notwithstanding, shorter positions seemed to be more profitable when examining fees earned per unit of time; however, the fees on such short timeframes were overshadowed by their IL. The shortest-lived positions (of those longer than one hour) generally and consistently had the worst fee-over-IL (akin to risk-adjusted returns) ratios, and the longest positions (over 1 month) had the most forgiving fee-over-IL ratios, although both groups lost money. To explain this effect, we refer to our earlier economic analyses of the AMM model, and specifically the fact that IL grows in square root time while fees accumulate linearly in time. While compelling, this explanation is still incomplete, and the time-dependence of the fee-to-IL ratios will require further examination.

The most fertile area for further research that we have identified is to refine our LP segmentation, and to explicitly identify the users that that have positive returns after IL, ie those that outperform the HODL strategy. This is particularly interesting because the protocol on average does worse than HODL, so every marginal dollar won by a fortunate Uniswap LP is a marginal dollar lost by a less fortunate one.

# APPENDIX

IL calculation in Uniswap v3

To value a user's activities on Uniswap V3, the number of each token associated with their liquidity position must be interrogated. This in non-trivial, as their quantities at any point in time are under the influence of variables that are not recorded to the blockchain. For example, the tick number, and therefore the effective price at the same moment a user adjusts their position is not readable. This is problematic as it is implicitly required to calculate the user's remaining balances. However, these values can be deduced using data that is available from the blockchain.

During an adjustment to a user's liquidity position – whether it be a withdrawal, or contribution of additional liquidity – the relative quantity of both tokens associated with the change must be equal to the relative quantity of tokens present inside the position before, and after the change is made. For example, it may be observed that a user added additional tokens to their Uv3 position, where 1×TKN0 and 2×TKN1 were transferred into an existing position. The ratio of TKN0 relative to the total token count, 1/(1 + 2), denoted here as *r*, is identical to their ratio inside the existing position. Combined with the variables that are available from the blockchain (tick ranges and the liquidity constant), the precise balance of TKN0 can be calculated from the ratio of the two assets (*r*), and the prices at the corresponding tick boundaries ($P_b$ and $P_b$):

$$x = \frac{L\left(\sqrt{P_a}\sqrt{P_b}(r-1) - r + \sqrt{P_a P_b (r-1)^2 + r^2 + (2r-2)\left(\sqrt{P_a}\sqrt{P_b}r - 2P_b r\right)}\right)}{2P_b r}$$

With the quantity *x* in hand, the quantity of TKN1 (*y*) can be calculated directly from *x* via the Uv3 swap function:

$$y = \frac{L\left(-L\sqrt{P_a} + L\sqrt{P_b} - \sqrt{P_a}\sqrt{P_b}x\right)}{L + \sqrt{P_b}x}$$

Uv3's internal price ($P_i$) can also be calculated directly from *x*:

$$P_i = \frac{L\sqrt{P_a}\sqrt{P_b}}{L + \sqrt{P_b}x} + \frac{L\sqrt{P_b}\left(-L\sqrt{P_a} + L\sqrt{P_b} - \sqrt{P_a}\sqrt{P_b}x\right)}{\left(L + \sqrt{P_b}x\right)^2}$$

And the tick ($T_i$) can be calculated directly from the price ($P_i$):

$$T_i = \frac{\log(P_i)}{\log(1.0001)}$$

To compute the impermanent loss, we use a novation method that simulates the complete withdrawal of all liquidity from the position, followed by a simulated redeposit, thus allowing the partial IL to be realized over an arbitrary number of add/remove liquidity interactions via the Uv3 contracts. To achieve this, the above equations are solved at two different time points to determine $x$ and $y$, using the appropriate $r$ value as observed on the blockchain, and the same liquidity constant, $L$, for each. The initial and final quantities of TKN0 and TKN1 are denoted as $x_0$ and $x_1$, and $y_0$ and $y_1$, respectively. Of course, only the price associated with the final time point, $P_1$, is required to determine the IL over this period:

$$IL = \frac{P_1 x_1 + y_1}{P_1 x_0 + y_0} - 1$$

Glossary:

$L$ is the liquidity constant

$P_a$ is the lower price (upper tick)

$P_b$ is the upper price (lower tick)

$x$ is the TKN0 amount

$y$ is the TKN1 amount

$r$ is the ratio between the two assets as defined by $y/(x+y)$

$P_i$ is the price

$T_i$ is the tick corresponding to the price ($P_i$)

Data sources

1. Transaction details of the NFT mint are obtained from *uniswap_v3."NonfungibleTokenPositionManager_call_mint"* on Dune.
2. Increase and decrease liquidity events are obtained from *uniswap_v3."NonfungibleTokenPositionManager_evt_IncreaseLiquidity"* and *uniswap_v3."NonfungibleTokenPositionManager_evt_DecreaseLiquidity"*, respectively on Dune.
3. Burnt tokens are obtained from *uniswap_v3."NonfungibleTokenPositionManager_call_burn"* on Dune.
4. Fee collection events are obtained from *uniswap_v3."NonfungibleTokenPositionManager_evt_Collect"* on Dune.
5. Hourly USD price data for token0 and token1 were obtained from *prices.usd* on Dune noting that not all tokens are available from this query.
6. Alchemy Queries are used to obtain current state variables:
   - Current *liquidity*, *feeGrowthInside0LastX128*, *feeGrowthInside1LastX128* are obtained from the *Uniswap V3: Positions NFT (UNI-V3-POS)* contract (0xC36442b4a4522E871399CD717aBDD847Ab11FE88).
   - *sqrtPriceX96* and current *tick* are obtained from the specific pool contract (*slot0* function).
   - *feeGrowthGlobal0X128* and *feeGrowthGlobal1X128* are obtained from the specific pool contract (*feeGrowthGlobal0X128*, *feeGrowthGlobal1X128* functions respectively).
   - *feeGrowthOutside0X128_upper*, *feeGrowthOutside1X128_upper*, *feeGrowthOutside0X128_lower*, *feeGrowthOutside1X128_lower* are obtained from the specific pool contract (*ticks* function, for each upper and lower tick, for token0 and token1 respectively).
7. External trade volume and fees check were obtained from *DEX.trades* query on Dune.

Technical Considerations

Overall, this analysis was extremely interesting, but also surprisingly complex. Firstly, the concept of IL is not easy to define. In particular, the numeraire dependence can never be avoided when handling data that includes funds being contributed or withdrawn over time. While IL can be expressed as a dimensionless percentage number, this number is not invariant under numeraire transformations. This is inconvenient, and annoying, but ultimately unavoidable. It should be noted that while the numbers are not invariant, the variance does not affect the result substantially when considering numeraires that have been evolving largely in line. It becomes important only when the different numeraires diverge dramatically (say by a factor of 3x or more) because in this case it impacts how past IL is carried forward in time. So, while our numbers are not invariant under numeraire change, the main finding – Impermanent Loss exceeds fees earned and therefore the average Uniswap v3 LP would have been better off HODLing – remains robust.

Secondly, the technical complexity of the analysis was surprisingly high. This was partially driven by having to deal with many different pools, many different assets, and therefore many different exchange rates that need to be kept track of in a consistent manner. Also, the data that can be obtained is optimized for gas efficiency, and not for ease of access and there are some interesting tricks used within the Uniswap v3 code, eg, unsigned integers representing negative numbers in 2s complement meaning that the formulas must be evaluated mod at every step when running in Python that does not allow for fixed length integers. For transparency, we offer our methodology and dataset from whence this conclusion was derived, to the DeFi community for review and scrutinization.